\documentclass{aa}  

\usepackage{graphicx}
\usepackage{txfonts}
\usepackage{lipsum}
\usepackage{subcaption} 
\usepackage{bm}

\usepackage{lscape}             
                               
\usepackage{placeins}

\usepackage[colorlinks=true, linkcolor=blue, citecolor=blue, urlcolor=blue]{hyperref}

\begin{document}

   \title{Stability mapping of the new Uranian moon S/2025 U1 with updated masses for Cordelia, Ophelia, and Cressida}
   \subtitle{}

   \author{J. Xavier\inst{1,2,3}\fnmsep\thanks{jadilene.xavier@ifal.edu.br}
        \and A. Amarante\inst{2,3}\fnmsep{}
        \and S. Giuliatti~Winter \inst{3} \fnmsep{}
           \and A. Ferreira \inst{2,3} \fnmsep{} }

   \institute{Federal Institute of Education, Science and Technology of Alagoas (IFAL), Palmeira dos Índios, Alagoas, Brazil
            \and Group of Orbital \& Rotational Research in Irregular Objects \& Observational astroNomy (ORION), Unesp, Brazil
            \and Group of Orbital Dynamics \& Planetology, São Paulo State University (Unesp), School of Engineering and Sciences, Guaratinguet\'a, S\~ao Paulo, Brazil \\}

   \date{Received September 30, 20XX}

  \abstract{
We investigate the dynamics of Uranus's inner satellite system considering three recent updates: the inclusion of the new moon S/2025~U1; revised mass estimates for Cordelia, Ophelia, and Cressida; and updated zonal harmonic coefficients (J$_2$, J$_4$, and J$_6$). Using numerical integrations, mean-motion resonance analysis, and frequency map analysis (FMA), we explored their impact on orbital stability.
For the first time, we analyzed the dynamics of the newly discovered moon S/2025~U1 and find that it follows a stable orbit, exhibiting smooth variations in its orbital elements over 250,000~years. Depending on the adopted radius, the gravitational influence of S/2025~U1 on the surrounding region is more compatible with a body of approximately 5-7 km in radius, producing diffusion maps associated with a regular orbital evolution. Furthermore, the case $R = 7$ km shows a reduction in the diffusion of neighboring satellites, suggesting a possible local stabilizing effect on the system's orbital evolution. We also identified inner and outer regions of low diffusion around its orbit, suggesting that the dynamical environment is compatible with the survival of coorbital particles or additional small satellites. In contrast, very high radius values, particularly $R = 20$ km, tend to increase diffusion among neighboring satellites, making this scenario less compatible with a dynamically stable orbital configuration. We further explored the survival of hypothetical satellites near S/2025~U1. Our results show that bodies with masses up to $3\times$ that of S/2025~U1 can remain stable in both the interior (between S/2025~U1 and Ophelia) and the exterior (between S/2025~U1 and Bianca) of its orbit.
Finally, we find that the group formed by Cressida, Desdemona, Juliet, and Portia is the most sensitive to the updated parameters, showing noticeable changes in orbital stability and in the evolution of maximum eccentricity. While most mean-motion resonances among the satellites exhibit a circulating behavior, the 44:43 resonance between Belinda and Perdita acts as a confinement mechanism for the Belinda-Cupid-Perdita system, with a libration amplitude smaller than previously reported.

}
\titlerunning{Stability mapping of the new Uranian moon S/2025 U1}
\authorrunning{J. Xavier et al. (2026)}
   \maketitle
\nolinenumbers 

\section{Introduction}
Uranus's inner satellites, discovered mainly by Voyager~2 in 1986, form a group quite distinct from other satellite systems in the Solar System. They are small and dark, and they follow nearly circular orbits very close to Uranus's equatorial plane \citep{Jacobson1998}. Furthermore, these satellites exhibit a chaotic, highly dynamic ``orbital dance'' \citep{Duncan1997}. The group includes Cordelia, Ophelia, Bianca, Cressida, Desdemona, Juliet, Portia, Rosalind, Cupid, Belinda, Perdita, Puck, and Mab, some of which were discovered after the Voyager~2 flyby. 

A subgroup of these satellites is known as the Portia family (Bianca, Cressida, Desdemona, Juliet, Portia, Rosalind, Belinda, and Puck). These moons orbit very close to one another, at distances too small to ensure long-term dynamical stability. Recent studies have shown that due to their proximity, some of these moons may cross orbits within a few million years, leading to chaotic gravitational interactions  \citep{French2012}.

Mab, discovered in 2003 by \citet{Showalter2006}, presents a peculiar characteristic. It shares its orbit with the tenuous and dusty $\mu$~ring. Recent studies suggest that the Mab $\mu$~ring system may be significantly influenced by small satellites undergoing horseshoe encounters, whose occasional collisions with Mab result in a dynamic, unstable orbital configuration, with evidence of accretion processes on timescales of approximately $10^6$ to $10^8$~years \citep{Kumar2015}.

Located very close to Uranus's surface, the small satellites Cordelia and Ophelia act as shepherd satellites, contributing to the confinement of the $\varepsilon$~ring and to its precession rate. Recent studies also indicate that these satellites are in resonance with other rings distinct from the $\varepsilon$~ring, intensifying the complex dynamics of the region where they orbit \citep{French2024}.

Cupid is part of one of the most complex groups of satellites, the Cupid-Belinda-Perdita group. This group is one of the most influential in the unstable orbital configuration of the system. Studies suggested that Cupid and Belinda could cross their orbits within an integration time of $10^3$ to $10^7$~years \citep{French2012}. However, more recent research using updated orbital data indicates that the satellites Belinda, Perdita, and Cupid may belong to a stable region that may persist for up to $10^8$~years, due to a mean-motion resonance between Belinda and Perdita \citep{Cuk2022}.

Recently, a new moon was identified within this intriguing group of satellites. Provisionally designated S/2025~U1, it is located in the orbital region between Ophelia and Bianca. The available information is still limited, but preliminary data indicate that its orbit is nearly circular and that the body has an estimated radius of approximately 5~km \citep{Maryame2025}.

The small orbital spacing among Uranus's inner satellites along with the proximity and likelihood of numerous resonances between the satellites themselves and between satellites and rings combined with uncertainties in mass estimates and other orbital parameters make the inner Uranian system particularly relevant for dynamical studies. In recent years, works have significantly refined the orbital data of these satellites \citep{Jacobson1998,Hedman2016,Cuk2020,Duncan1997,Quillen2014,French2012,French2015,Chancia2017}. However, uncertainties in mass determinations persist. In a recent study, \citet{French2024} used the gravitational perturbations observed in the rings to update the masses of the satellites Cressida, Cordelia, and Ophelia as well as to provide revised values for Uranus's zonal harmonic coefficients J$_2$, J$_4$, and J$_6$. The values of Uranus's gravitational coefficients were also updated in a recent work \citep{Jacobson2025}.

In light of the new data presented on Uranus's inner satellites and the growing interest in future missions to the planet, a reassessment of the stability of this system has become necessary \citep{Lorenzo2023,Girija2023,Fletcher2020}. The objective of this study is to evaluate the impact of recent updates to the Uranian inner satellite system on its dynamical behavior and to investigate the dynamical structure and stability of the region surrounding the newly discovered moon.

This article is organized as follows. In Sect. \ref{sec:Disc} we present a detailed discussion of the main results obtained throughout this study. We used the frequency map analysis (FMA) technique \citep{Laskar1990,Laskar1993} to investigate the diffusion of the orbits of Uranus's inner satellites, incorporating, for the first time, the updated mass values of Cordelia, Ophelia, and Cressida \citep{French2024} as well as the recently identified moon S/2025~U1 \citep{Maryame2025}. Additionally, we included the revised values of the gravitational coefficients J$_2$, J$_4$, and J$_6$ \citep{Jacobson2025}. The mean-motion resonances analyzed between the satellites and the test particles are described in Sect. \ref{sec:res}. The dynamics of the newly identified moon S/2025~U1 and its surrounding region are examined in Sect. \ref{sec:new_moon}. In Sect. \ref{sec:hypsat} the dynamical viability of additional satellites in the inner and outer regions surrounding the recently discovered moon S/2025~U1 is investigated through simulations that consider the presence of hypothetical satellites. Finally, in Sect. \ref{sec:concl}, we summarize the main conclusions.

\section{Frequency map analysis and numerical simulations}
\label{sec:FMA_NS}
\begin{table*}
\centering
 \caption{Data from Uranus's inner satellites.}
\label{tab_1}
\scalebox{0.57}{
\begin{tabular}{@{}lllllll@{}}
\hline
\hline
    Satellite    & $a$~(km)                      &  $e$                     & $I$ ($^\circ$)        & $\omega$($^\circ$)          & $\Omega$($^\circ$)           & $M$($^\circ$)           \\ \hline
    Cordelia    &  49,752.000 \citet{French2024}  &  0.0003 \citet{Cuk2022}   & 0.085 \citet{Cuk2022}  & 136.82711 \citet{Jacobson1998}                    & 38.37431  \citet{Jacobson1998}          & 254.80512 \citet{Jacobson1998}                    \\
    Ophelia     & 53,764.000 \citet{French2024}   & 0.00948 \citet{Cuk2022}  & 0.09080 \citet{Cuk2022} & 183.58541 \citet{Cuk2022}    & 87.00642 \citet{Cuk2022}      & 321.39992 \citet{Cuk2022}  \\
     Bianca     &  59,165.000 \citet{French2024}  & 0.00027 \citet{Cuk2022}  & 0.1811  \citet{Cuk2022} & 11.002 \citet{Showalter2006} & 288.141 \citet{Showalter2006} & 20.337 \citet{Showalter2006}  \\
     Cressida   &  61,767.000 \citet{French2024}  & 0.00056 \citet{Cuk2022}   & 0.05023 \citet{Cuk2022} & 67.20697 \citet{Cuk2022}    & 207.44872 \citet{Cuk2022}     & 223.04548 \citet{Cuk2022}   \\
     Desdemona  & 62,659.000 \citet{French2024}   & 0.00073 \citet{Cuk2022}   & 0.04673 \citet{Cuk2022} & 35.50588 \citet{Cuk2022}    & 122.42866 \citet{Cuk2022}     & 194.96438  \citet{Cuk2022}   \\
     Juliet     & 64,358.000 \citet{French2024}   & 0.00122 \citet{Cuk2022}   & 0.03518 \citet{Cuk2022} & 305.68199 \citet{Cuk2022}   & 144.28899 \citet{Cuk2022}     &346.20368  \citet{Cuk2022}   \\
     Portia     & 66,097.000 \citet{French2024}   & 0.00051 \citet{Cuk2022}   & 0.01643 \citet{Cuk2022} & 335.16597 \citet{Cuk2022}   & 31.75335 \citet{Cuk2022}      & 43.52059 \citet{Cuk2022}     \\
     Rosalind   & 69,927.000 \citet{French2024}   & 0.00090 \citet{Cuk2022}   & 0.04922 \citet{Cuk2022} & 350.32405 \citet{Cuk2022}    & 249.40149 \citet{Cuk2022}     & 260.12628 \citet{Cuk2022}    \\
     Cupid      & 74,392.338 \citet{Cuk2022}      & 0.0004 \citet{Cuk2022}    & 0.07028 \citet{Cuk2022} & 260.48641 \citet{Cuk2022}   & 59.71798 \citet{Cuk2022}      & 148.86401 \citet{Cuk2022}   \\
     Belinda    & 75,255.000 \citet{French2024}   & 0.00079 \citet{Cuk2022}   & 0.00172 \citet{Cuk2022} & 166.49423 \citet{Cuk2022}  & 232.28659 \citet{Cuk2022}     & 285.8812 \citet{Cuk2022}     \\
     Perdita    & 76,416.764 \citet{Cuk2022}      & 0.00345 \citet{Cuk2022}    & 0.05889\citet{Cuk2022} & 90.14331  \citet{Cuk2022}   & 248.66870 \citet{Cuk2022}     & 88.67752 \citet{Cuk2022}   \\
     Puck       & 86,004.000 \citet{French2024}   & 0.00010 \citet{Cuk2022}   & 0.33114 \citet{Cuk2022} & 261.39828 \citet{Cuk2022}   & 235.33913 \citet{Cuk2022}     & 169.70695 \citet{Cuk2022}   \\
     Mab        & 97,735.966 \citet{Cuk2022}      & 0.00347 \citet{Cuk2022}    & 0.12217\citet{Cuk2022} & 65.29087  \citet{Cuk2022}   & 84.63121 \citet{Cuk2022}      & 3.14271 \citet{Cuk2022}     \\ \hline
    \end{tabular}}
\end{table*}
\begin{table}
\centering
  \caption{Parameters of Uranus.}
\label{tab_2}
\scalebox{1.4}{
\begin{tabular}{@{}ll@{}}
\hline
\hline
 \textsuperscript{1} Parameter & Value \\ \hline
     J$_2$     & $3508.967 \times 10^{-6}$  \\
      J$_4$     & $-35.793 \times 10^{-6}$  \\
      J$_6$    & $0.575 \times 10^{-6}$  \\
      $R_U$       & 25,559 (km)      \\ \hline
    \end{tabular}}
    \textsuperscript{1}{\begin{minipage}[t]{\linewidth} \citet{Jacobson2025}.
    \end{minipage}}
   \end{table}
Dynamic maps are widely used to investigate regions of orbital stability and instability for bodies in the Solar System. This tool has been employed, for instance, in the study of Uranus's inner satellites \citep{Charalambous2022}. In that work, the authors analyzed the complex dynamics among moons such as Cupid, Belinda, Perdita, and Rosalind, identifying, through N-body simulations and dynamic maps, an intricate web of mean-motion resonances among these satellites. The study also suggests that interactions with a circumplanetary disk, as well as tidal effects with Uranus, may have driven orbital migrations and resonance captures, shaping the current configuration of the system.

\subsection{Frequency map}
These maps can be constructed using the FMA, initially introduced by \citet{Laskar1990}. Based on the Fourier transform, this method allows exploration of the system's global behavior through short-term, computationally inexpensive numerical integrations, making it particularly suitable for examining large regions of the system's phase space.

In \citet{Laskar2001}, the dynamics of massless particles in the Solar System is analyzed using the FMA technique. The domain studied ranges from the orbit of Mercury (0.38\,au) to the outer regions of the Kuiper Belt (90~au), to identify zones of orbital stability and instability. The application of the method provided a detailed understanding of the dynamical evolution of these particles, revealing the complex interplay between regular and chaotic regions. The results, for instance, highlighted intricate dynamical structures within the asteroid belt, strongly influenced by resonances with Jupiter and Saturn.

The chaotic behavior of comet Halley's orbit was also investigated in the work of \citet{Munoz-Gutierrez2015}. The authors use the FMA to identify regions on the $a\times e$ map corresponding to the most unstable orbits. The sensitivity of the dynamics to initial conditions near Halley's orbit indicates that, on average, it is unstable on a time scale of hundreds of thousands of years.

The FMA is also employed in the work of \citet{Gutierrez2017} to investigate the long-term orbital evolution and dynamical stability of four small Saturnian satellites: Aegaeon, Methone, Anthe, and Pallene. The study concluded that Aegaeon, Methone, Anthe, and Pallene exhibit stable orbits over an integration time of up to 100,000~years. Resonances with Mimas play a crucial role in maintaining this stability, particularly for Aegaeon, Methone, and Anthe. Pallene, although not in a direct resonance, is influenced by a quasi-resonance that contributes to its orbital stability.

A similar study was conducted by \citet{Gaslac2020}, who investigated the dynamics of the region encompassing Neptune's rings and its small inner satellites using diffusion maps. The study concludes that Neptune's inner rings, particularly the Galle ring, are located in dynamically stable regions. In contrast, the Lassell, Le Verrier, and Adams rings can remain stable as long as their eccentricities remain below 0.012. The satellites Naiad, Thalassa, and Despina are identified as potential sources of particle replenishment for the rings, unlike Galatea, Larissa, and Proteus, whose ejecta remain gravitationally bound. Using diffusion maps, the authors also identified zones of stability and resonant regions, further enhancing understanding of the interactions between the rings and Neptune's small inner satellites.

In this work, the construction of dynamic maps was carried out using the Frequency Modified Fourier Transform (FMFT) algorithm, developed by \citet{Nesvorny1996}, to estimate the dominant frequency $\nu_0$ of a complex variable defined as $z' (t) = a(t) e^{i\lambda(t)}$, where $a$ corresponds to the semimajor axis and $\lambda$ to the mean longitude of a test planet. Both quantities are time-dependent and obtained from a short numerical integration. Following the approach proposed by \citet{Laskar2001}, the dominant frequency derived from the complex variable \( z' (t) \) is strongly associated with the mean-motion \( n \) of the orbiting body. Within the framework of frequency analysis, this time-dependent function can be represented as a finite sum of exponential components:
\begin{equation}
z'(t) = \alpha_0 e^{i \nu_0 t} + \sum_{j=1}^{N} \alpha_j e^{i \nu_j t},
\end{equation}
\noindent where $\nu_0$ denotes the primary frequency component. In a purely Keplerian scenario, $\nu_0$ would coincide with the mean-motion $n$. However, due to dynamical perturbations, deviations from this ideal are expected, and both the amplitude $|\alpha_0|$ and the frequency $ \nu_0$ remain close to, but not precisely equal to, the values associated with the initial semimajor axis and corresponding mean-motion.

To describe orbital stability, we introduce a numerical indicator that quantifies the variation in the dominant orbital frequency over a given time interval. This parameter, referred to as the diffusion index $D$, measures the difference between the average orbital frequencies computed over two distinct time intervals. It is defined as $ D = |\nu_{01} - \nu_{02}|/T$ \citep{Gaslac2024}. Where $\nu_{01}$ and $\nu_{02}$ represent the mean frequencies calculated over the first and second halves of the total integration time $2T$, respectively, and $T$ corresponds to the duration of each interval. This metric provides a quantitative estimate of orbital stability: low values of $D$ indicate that the frequency remains nearly constant, suggesting regular motion, whereas high values reveal significant variations, which may indicate instability or even chaotic behavior over a time interval of $2T$.

\subsection{Mathematical model and numerical simulations}
\label{sec:model}
The system under study consists of Uranus, acting as the central body, and its thirteen inner satellites orbiting around it. To model the orbital dynamics of Uranus's satellites, we adopt an inertial reference frame centered at the planet's center of mass. In this context, the dominant gravitational force acting on the inner satellites is Uranus's own gravity field. This field can be described through a spherical harmonics expansion, as introduced by \citet{Kaula1966}, that allows the gravitational potential \( U \) to be expressed as an infinite series:
\begin{small} 
\begin{equation}
\begin{aligned}
U(r,\theta,\lambda) &= \frac{G M_U}{r}
\Bigg\{ 1 + \sum_{\ell=2}^{\infty} \sum_{m=0}^{\ell}
\left( \frac{R_U}{r} \right)^{\ell} \\
&\qquad \times
\big[ C_{\ell m}\, Y_{\ell m1}(\theta,\lambda)
     + S_{\ell m}\, Y_{\ell m0}(\theta,\lambda) \big]
\Bigg\},
\end{aligned}
\end{equation}
\end{small}
\noindent where $GM_U$ is Uranus's gravitational parameter and $R_U$ is the equatorial radius of Uranus, $r$ is the radial distance of the satellite from the planet's center, and $\theta$ and $\lambda$ are the satellite's colatitude and longitude, respectively. The functions $Y_{\ell m}$ denote the associated spherical harmonics. The coefficients $C_{\ell m}$ and $S_{\ell m}$ represent the mass distribution of Uranus. In this reference frame, the coefficients $C_{10}$, $C_{11}$, and $S_{11}$ vanish by definition, as the origin is placed at the planet's center of mass.

We truncated the expansion at $\ell = 6$, which allowed us to retain the main zonal coefficients of Uranus's gravity field. It includes Uranian zonal harmonic coefficients J$_2$, J$_4$, and J$_6$, which characterize, respectively, the planet's equatorial flattening and subtler deviations from spherical symmetry in its internal mass distribution. These terms are essential for accurately capturing the effects of Uranus's nonspherical gravity field on the orbital motion of its nearby satellites \citep{Xavier2023}.

We employed geometric orbital elements, following the methodology proposed by \citet{Borderies1994} and later implemented numerically by \citet{Sicardy2006}. This approach correctly converts orbital elements into state vectors and vice versa, accounting for the Uranian zonal harmonic coefficients J$_2$, J$_4$, and J$_6$ in numerical simulations.

Unlike osculating elements, which describe an instantaneous Keplerian orbit tangent to the satellite's actual trajectory and are therefore subject to rapid oscillations caused by perturbations such as Uranian zonal harmonic coefficients (J$_2$, J$_4$, and J$_6$), geometric elements correspond to a secular average of the orbit, effectively filtering out short-period variations. This makes them particularly suitable for long-term orbital evolution studies, including secular precession, resonance analysis, and dynamical stability \citep{Callegari2020}. Consequently, the geometric elements provide a more precise description of the orbital architecture of Uranus's inner satellites.

We performed numerical integrations of the Uranus inner moons system over $10^8$ orbital periods of its outermost moon, Mab, and, for the construction of FMA with 10,000 particles, over $10^4$ periods of Mab. 
Frequency analysis is particularly efficient for investigating the orbital stability of planetary systems with short integration times \citep{Gaslac2020}. Here, the novelty lies in using the FMA to infer the long-term stability of a highly compact system, such as the Uranian inner satellites, at significantly reduced computational cost.

The numerical simulations were conducted in three main stages: (i) in the first stage, we used the orbital parameters listed in Table~\ref{tab_1} and the standard mass values provided in Table~\ref{tab_3}; (ii) in the second stage, we adopted the updated mass values for Cordelia, Cressida, and Ophelia proposed by \citet{French2024} and also listed in Table~\ref{tab_3}; (iii) in the 3$^{rd}$ stage, we employed the new mass values for Cordelia, Ophelia, and Cressida and additionally incorporated the newly identified satellite S/2025~U1 into the system. The purpose of including this new moon is to analyze its influence on the system's global dynamics. Preliminary results suggest that S/2025~U1 has a semimajor axis of approximately 56,000 km and a nearly circular eccentricity. Further data on its orbit have not yet been provided. The parameters associated with this satellite used in this work are presented in Table~\ref{tab:new_moon}. 

Additionally, we employed updated values for Uranus's zonal harmonic coefficients based on the most recent data \citep{Jacobson2025}, as shown in Table~\ref{tab_2}. The integrations were carried out using the Bulirsch Stoer integrator from the MERCURY N-body package \citep{Chambers1999}.

To analyze the evolution and stability across the region encompassing the 14 satellites incorporating the updated mass values for Cordelia, Cressida, and Ophelia, as well as the newly identified moon S/2025~U1, frequency analysis maps were constructed to calculate the diffusion parameter from the output files of the MERCURY package with the geometric element approach \citep{Borderies1994,Sicardy2006}.

\section{Analysis and discussion of the results}
\label{sec:Disc}
\begin{table}
\centering
  \caption{Masses of Uranus's satellites corresponding to different sets of simulations.}
\label{tab_3}
\scalebox{0.7}{
\begin{tabular}{@{}lll@{}}
\hline
\hline
Satellite & Standard mass (kg)
 & New mass (kg)\\ \hline
  \textsuperscript{2}   Cordelia     & $3.05 \times 10^{16}$ \citet{French2024} & $6.08 \times 10^{16}$ \citet{French2024}\\
   \textsuperscript{2}   Ophelia     & $3.67 \times 10^{16}$ \citet{French2024} & $3.57 \times 10^{16}$ \citet{French2024} \\
      Bianca      & $6.38 \times 10^{16}$ \citet{French2024} & Standard mass \\
  \textsuperscript{2}    Cressida    & $23.74 \times 10^{16}$ \citet{French2024} & $18.39 \times 10^{16}$ \citet{French2024} \\
      Desdemona    & $12.37 \times 10^{16}$ \citet{French2024} & Standard mass\\
    Juliet        & $38.71 \times 10^{16}$ \citet{French2024} & Standard mass\\
    Portia        & $116.71 \times 10^{16}$ \citet{French2024} & Standard mass\\
    Rosalind      & $17.59 \times 10^{16}$ \citet{French2024}     &     Standard mass \\
    Cupid         & $0.295297 \times 10^{16}$ \citet{French2015}    &    Standard mass \\
    Belinda       & $24.71 \times 10^{16}$ \citet{French2024}     & Standard mass \\
    Perdita       &  $0.985470 \times 10^{16}$ \citet{French2015}    &  Standard mass \\
    Puck          & $200.35 \times 10^{16}$ \citet{French2024}    &  Standard mass \\
    Mab           & $0.79865 \times 10^{16}$ \citet{French2015}    &  Standard mass \\ \hline
 \end{tabular}}
\tablefoot{\textsuperscript{2} For the simulations using the new mass values, we adopted the following densities: 1.79~g/cm³, 0.87~g/cm³, and 0.70~g/cm³ for Cordelia, Ophelia, and Cressida, respectively. For the simulations with the standard mass values, we used a density of 0.90~g/cm³ for all satellites. Both sets of density values were taken from the work of \citet{French2024}.}

\end{table}
\begin{table}
\centering
  \caption{Preliminary parameters of the new moon S/2025~U1.}
\label{tab:new_moon}
\scalebox{1.15}{
\begin{tabular}{@{}ll@{}}
\hline
\hline
 \textsuperscript{1} Parameter & Value \\ \hline
     $a$     & $\sim 5.6 \times 10^{4}$ (km) \\
      $e$     &  $\sim 0$   \\
      $R$    & $\sim 5.0$ (km) \\ \hline
  
    \end{tabular}}
    \textsuperscript{1}{\begin{minipage}[t]{\linewidth}\citet{Maryame2025}.
    \end{minipage}}
   \end{table}
\begin{figure*}
\includegraphics[width=\linewidth]{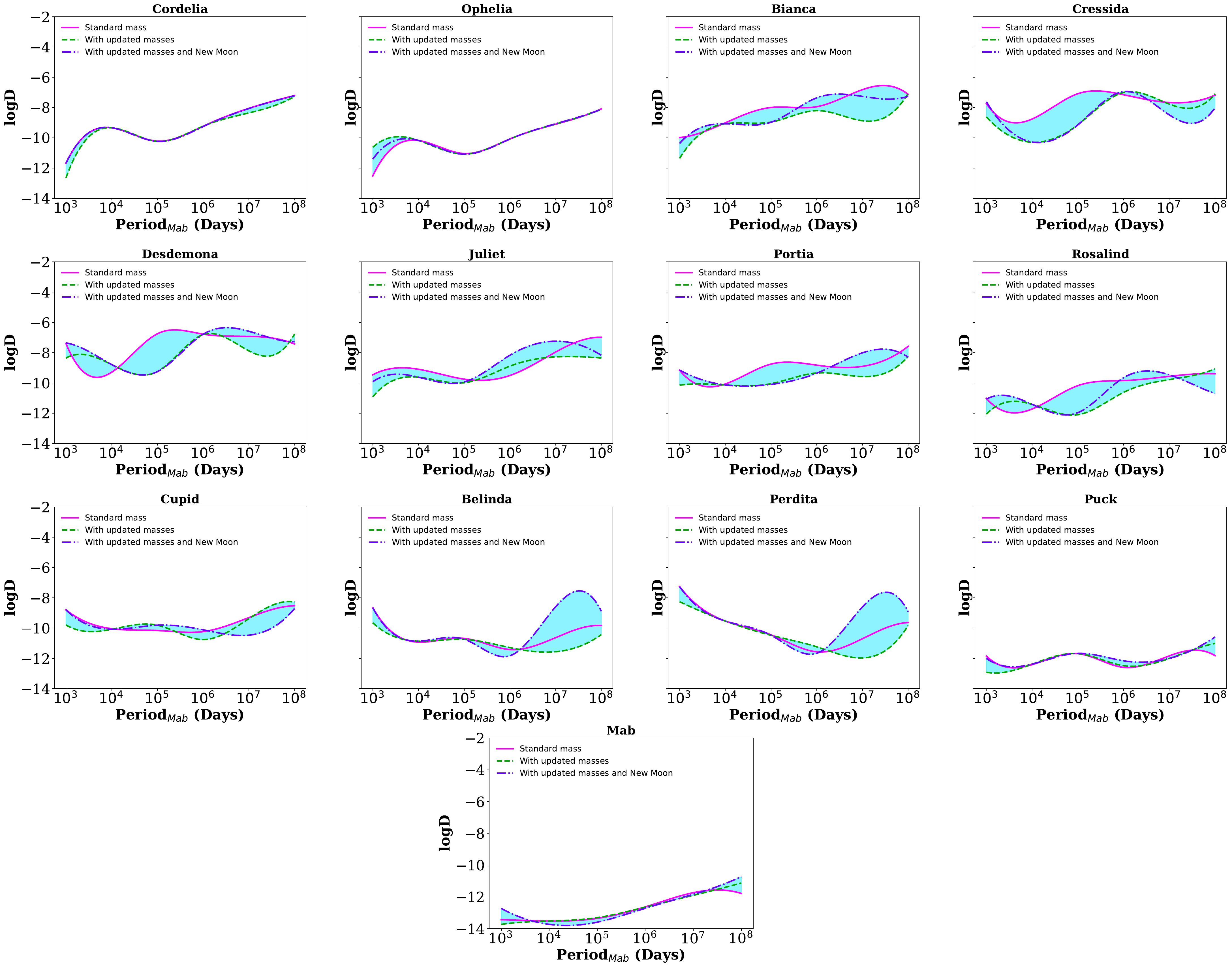}
 \caption{Diffusion parameter ($D$) computed through FMA as a function of the integration time (in Mab's orbital periods) for 13 inner satellites under three distinct scenarios: (i) using the standard masses (pink line); (ii) using the updated mass values for Cordelia, Cressida, and Ophelia (dotted green line), as presented in Table \ref{tab_3} (New Mass); and (iii) using the complete system with updated masses for Cordelia, Cressida, and Ophelia along with the inclusion of the newly identified moon S/2025~U1 (dotted~purple line). The horizontal axis represents the integration time (2T ) used to compute the orbital frequencies rather than the physical time evolution of the system. } 
\label{fig:log_DxTime}
\end{figure*}
Variations in the diffusion parameter ($D$) over time reflect changes in the dynamical structure of the Uranian inner satellite system under different physical and initial conditions. In particular, higher or moderate values of $D$ are associated with regions where orbital evolution is more sensitive to perturbations, while lower values indicate more regular motion \citep{Munoz-Gutierrez2015}. 

By comparing the diffusion behavior across different mass assumptions and system initial configurations, it is possible to identify how modifications to the physical and initial parameters affect the orbital stability of the system over a given time interval $2T$. The comparison of these different physical scenarios aims to evaluate the orbital dynamics relative to a more plausible scenario that represents the current dynamical behavior of the system.

To facilitate the interpretation of the diffusion analysis, a qualitative classification of the behavior of each satellite based on the results shown in Fig. \ref{fig:log_DxTime} is presented at the end of this section. The "sensitivity" classification describes the magnitude of the variations in the diffusion parameter ($\log D$) among the three investigated scenarios. Satellites classified as "low" exhibit only small variations, whereas those classified as "moderate" or "high" exhibit progressively larger differences in their diffusion behavior. The "consistency" classification indicates whether the overall trend of the diffusion parameter remains similar throughout the range of integration times considered. Satellites classified as "consistent" exhibit comparable diffusion trends in the three scenarios, whereas those classified as "inconsistent" show considerable changes.

\subsection{Orbital evolution and stability}
\label{subsec:orbevostab}
In Fig. \ref{fig:log_DxTime}, we present an analysis of how the estimated value of the diffusion parameter $D$ depends on the integration time ($2T$) used in the spectral frequency analysis, expressed in Mab's orbital periods, for the orbits of the 13 previously known inner satellites of Uranus, considering the three scenarios described in Sect. \ref{sec:model}.

The blue regions indicate the intervals in which the differences between the scenarios produced the most significant variations in the estimated values of $D$. The horizontal axis represents the integration time ($2T$) used to compute the orbital frequencies, rather than the system's physical time evolution. Accordingly, variations in $D$ along this axis reflect changes in the resolution and convergence of the spectral frequency analysis in a given time $2T$. 

We observed that in all scenarios except for Belinda and Perdita, $D$ increases with increasing integration time $2T$, suggesting that the orbital evolution of the satellites is becoming unstable, as found in previous works that conducted long-term numerical simulations \citep{French2015}. The case of Belinda and Perdita suggests a mechanism that stabilizes orbital evolution over time integration of $2T$, and this is discussed in more detail in Sect. \ref{sec:mmr:BelPer}.

Comparing the three different scenarios (lines), Cordelia exhibits a consistent behavior across the full range of integration times, with only minor deviations around $10^{3}$ Mab orbital periods and, more noticeably, between $10^{6}$ and $10^{8}$ periods, where the curve associated with the updated masses shows slightly lower values of $D$. This behavior indicates that Cordelia's dominant orbital frequency remains stable and is only weakly affected by the adopted mass variations. 

Ophelia shows a gradual convergence between the scenarios for integration times longer than $10^{3}$ Mab orbital periods, with only a slight increase in the diffusion values $D$ between $10^{3}$ and $10^{4}$ periods, when the updated masses are adopted. This behavior indicates that the spectral frequency estimated is only weakly affected by the adopted mass variations. The variations in $D$ are consistent with the modest change in Ophelia's mass and support a dominant orbital frequency across the explored integration times $2T$.

The satellites of the Bianca, Cressida, Desdemona, and Juliet group were the most affected by the adoption of the updated mass values for Cordelia, Ophelia, and Cressida, as well as by the inclusion of the new moon S/2025~U1. Over the entire range of analyzed orbital periods, variations in $D$ are observed, indicating an increased sensitivity of the orbital stability of these orbits to the modifications introduced in the physical parameters of the system. 

Bianca displays more pronounced differences between the scenarios in the interval between $10^{6}$ and $10^{8}$ Mab orbital periods, where the diffusion values $D$ of the complete system lie between those obtained in the other configurations. This behavior indicates that Bianca is among the satellites most sensitive to the modifications introduced in the system configuration, particularly at long integration times.

Cressida, whose mass is directly modified in the adopted scenarios, shows a slight decrease in $D$ at the beginning of the analysis, in $10^{3}$ orbital periods of Mab. After this time, $D$ gradually increases to $10^6$, corresponding to the interval in which the greatest differences between the scenarios are observed. For longer integration times, $D$ increases moderately between $10^{6}$ and $10^{7}$ periods, followed by a gradual decrease up to $10^{8}$ periods. The largest differences between the scenarios are concentrated between $10^{3}$ and $10^{6}$ Mab orbital periods, indicating that Cressida is among the satellites most sensitive to the adopted modifications in the system configuration.

Desdemona shows a diffusion behavior $D$ closely resembling that of Cressida, with the main difference occurring at short integration times, where the diffusion $D$ computed using the standard masses is slightly lower than in the other scenarios. This similarity suggests that both satellites respond in a comparable way to the adopted system modifications.

Juliet maintains low diffusion values of $D$ over the full range of integration time ($2T$), with only a slight increase of $D$ beyond $10^{5}$ Mab orbital periods and no significant differences between the scenarios. Portia shows behavior comparable to that of $D$, with minimal variation. In both cases, the diffusion parameter shows a weaker dependence on the adopted system configuration than that observed for satellites such as Bianca, Cressida, and Desdemona. This indicates that the modifications introduced in the updated models produce comparatively smaller changes in the diffusion of Juliet and Portia.

On the other hand, Rosalind shows a slight increase in $D$ beginning at $10^{4}$ Mab orbital periods across all investigated scenarios, which persists until the end of the integration. The similar behavior observed in all scenarios indicates that the diffusion evolution of Rosalind is only weakly affected by the adopted modifications.

Belinda, Cupid, and Perdita exhibit a distinct dynamical behavior associated with their orbital configuration and mutual gravitational interactions within Uranus's inner system. At short and intermediate integration times, the diffusion parameter $D$ decreases, consistent with a spectral frequency resolution. Cupid, in particular, maintains consistently low diffusion values, indicating limited sensitivity to perturbations from neighboring satellites, although $D$ increases with increasing integration time.

For Belinda and Perdita, the increase in $\log D$ should be interpreted with caution. This behavior appears only for this pair, which is involved in the 44:43 mean-motion resonance, and is not accompanied by a transition from the resonant argument of libration to circulation. As shown in Sect. ~\ref{sec:mmr:BelPer}, the resonance angle remains bounded throughout the entire integration interval. Therefore, the increase in $\log D$ does not indicate escape from resonance but suggests that the frequencies extracted by the FMA are more sensitive to the dynamical model adopted for this resonant pair. Overall, these results indicate that Belinda and Perdita constitute the most dynamically sensitive region of the inner satellite system of Uranus, while still preserving its resonant configuration on the investigated timescales.

The satellites Puck and Mab, the outermost members of Uranus's inner system, exhibit low diffusion $D$ values across the entire range of integration times analyzed. In addition, their diffusion curves remain very similar in all investigated scenarios, indicating that their long-term dynamics are only weakly affected by the adopted changes in the model.

Puck, the most massive satellite in the system, maintains nearly constant values of $D$ over the full interval considered, indicating a remarkably uniform behavior among the three adopted configurations.
Mab also exhibits low diffusion values at short and intermediate integration times, with only a slight increase in $D$ at the longest integration times considered. Despite this increase, the diffusion curves remain closely grouped, suggesting that the adopted modifications produce only a limited effect on their diffusion value behavior.

 In the plots shown in Fig.~\ref{fig:log_DxTime}, the largest differences between the investigated scenarios are concentrated in the region occupied by Bianca, Cressida, and Desdemona, whereas most of the remaining satellites exhibit only minor changes in the diffusion parameter $D$. Belinda and Perdita constitute a separate case, displaying a distinct long-term behavior that is not shared by the other satellites. These results suggest that the dynamical consequences of the updated masses and the inclusion of S/2025~U1 are localized within specific portions of Uranus's inner satellite system, rather than affecting all satellites equally. For clarity, a qualitative summary of the sensitivity and consistency of the diffusion behavior for each satellite is provided in Table~\ref{tab:diffusion_summary}. The consistency classification refers to the overall agreement of the diffusion trends among the different integration times considered.
\begin{table}[!h]
\centering
\caption{Qualitative summary of the diffusion behavior and consistency of the inner Uranian satellites.}
\label{tab:diffusion_summary}
\scalebox{1.0}{
\begin{tabular}{lcc}
\hline
\hline
Moon & Sensitivity & Consistency \\
\hline
Cordelia  & Low      & Consistent     \\
Ophelia   & Low      & Consistent     \\
Bianca    & High     & Not Consistent \\
Cressida  & High     & Not Consistent \\
Desdemona & High     & Not Consistent \\
Juliet    & Moderate & Consistent     \\
Portia    & Moderate & Consistent     \\
Rosalind  & Low      & Consistent     \\
Cupid     & Moderate & Consistent     \\
Belinda   & High     & Not Consistent \\
Perdita   & High     & Not Consistent \\
Puck      & Low      & Consistent     \\
Mab       & Low      & Consistent     \\
\hline
\end{tabular}}
\end{table}

\subsection{Maximum eccentricity of satellites}
\label{subsec:maxecc}

\begin{figure*}[ht]
\centering
\includegraphics[width=\linewidth]{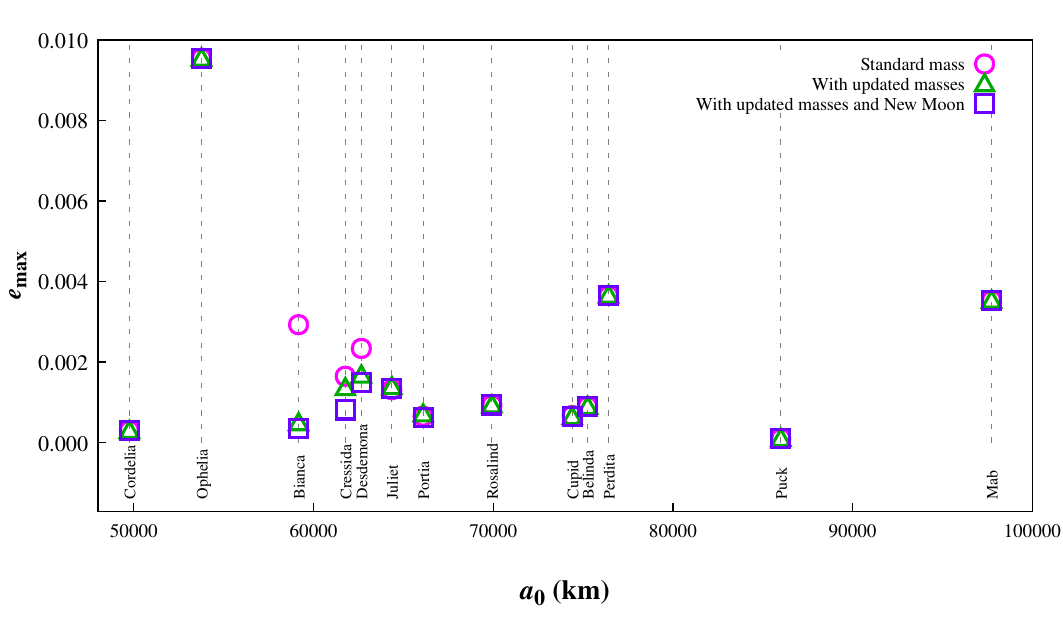}
\caption{Maximum eccentricity $e_{\max}$ as a function of the initial semimajor axis $a_0$ of the 13 inner satellites of Uranus under three distinct scenarios: (i) the standard masses (magenta circles); (ii) the updated mass values for Cordelia, Cressida, and Ophelia (green triangles); and (iii) the updated masses together with the inclusion of the newly identified moon S/2025~U1 (purple squares). The moons are directly identified along the horizontal axis.}
\label{fig:e_max}
\end{figure*}

To investigate changes in the orbital stability of satellites related to their orbital-element evolution, we analyzed the maximum eccentricity $e_{max}$ of the 13 inner satellites as a function of their initial semimajor axis $a_0$ over an integration period of \(10^4\) orbital periods of Mab. The results, presented in Fig.~\ref{fig:e_max}, consider three distinct scenarios: (i) standard masses (magenta circles); (ii) updated mass values for Cordelia, Cressida, and Ophelia (green triangles); and (iii) the complete system, which includes the updated masses along with the addition of the newly identified moon S/2025~U1 (purple squares).

Consistent with the analysis of the diffusion parameter $D$, the Bianca's group (formed by Bianca, Cressida, and Desdemona) exhibits the largest dynamical variation among the different scenarios, including the standard mass, the updated masses of Cordelia, Ophelia, and Cressida, and the complete system with the addition of the new moon. In particular, a decrease in $e_{max}$ is observed throughout the region extending approximately from $a \simeq 59,165$ km (Bianca's orbit) to $a \simeq 62,658$ km (Desdemona's orbit). 

Bianca exhibits the largest reduction in $e_{max}$ among all the satellites analyzed. In the standard mass scenario, $e_{max}=0.002929$, decreasing to 0.000496 and 0.000354 in the scenarios with updated masses and the complete system, respectively. This reduction by a factor greater than 8 indicates that Bianca's dynamics are particularly sensitive to modifications introduced in the dynamic model. It is possible to observe in Fig. \ref{fig:e_max} that Bianca's maximum eccentricity decreases significantly after the system updates, both due to the revision of the masses of some moons and the inclusion of S/2025~U1.

Cressida and Desdemona also exhibit a systematic reduction in $e_{max}$, although less pronounced. In the case of Cressida, the values change from 0.001650 to 0.001365 and 0.000813, while for Desdemona, the values vary from 0.002341 to 0.001678 and 0.001496. In both cases, the values remain within the same order of magnitude across the three scenarios considered, indicating that the updates affect their orbits to a lesser extent. This behavior is consistent with the results obtained for the diffusion parameter $D$, which also indicate the greatest dynamical differences in this region of the system.

This decrease in $e_{\max}$ suggests that the updated masses and the inclusion of the new moon affect the system's gravitational interactions, potentially leading to a temporary increase in orbital stability for the satellites' orbits in this region. In a highly packed system such as the inner Uranian satellites, small variations in the physical parameters can produce significant changes in the orbital evolution, directly affecting the amplitude of eccentricity variations. All values of $e_{max}$ for the analyzed scenarios are also listed in Table~\ref{tab:e_max}.

Therefore, the results of $e_{max}$ indicate that the dynamics of the Bianca group are particularly sensitive to the physical properties of the system. Modifications in the masses and orbital evolution can significantly alter the level of orbital stability, even without drastic changes in the overall initial configurations of the system.
\begin{figure*}
 \includegraphics[width=\textwidth]{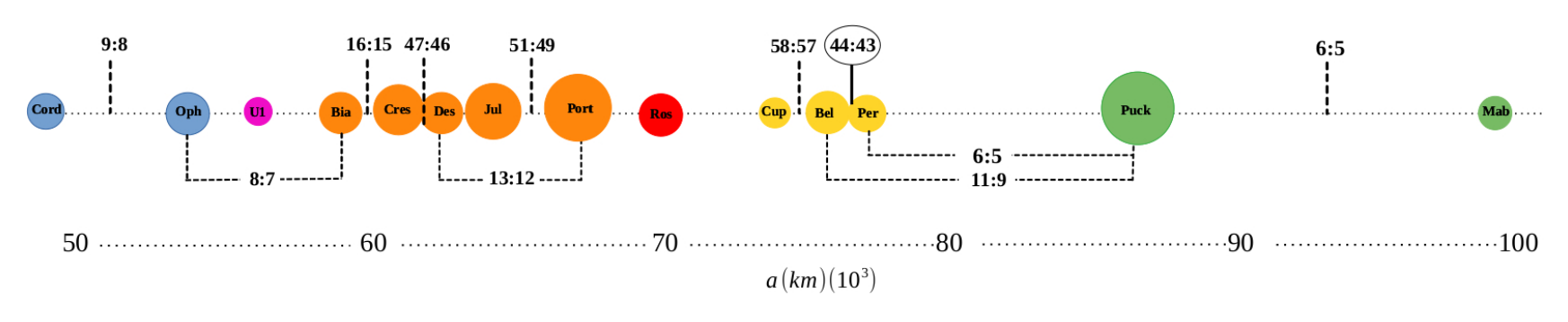}
        \caption{Representation of the main mean-motion resonances among Uranus's inner satellites. Some commensurability values are approximate and do not correspond to mean-motion resonances, indicated by the dotted lines, while resonances are shown with solid lines and highlighted with circles; see the discussion in Sect. \ref{sec:res}. The abbreviation U1 denotes the recently discovered moon S/2025~U1. Figure \protect is inspired by previous works \citep{French2015,Charalambous2022}.}
        \label{fig:All_Ress}
 \end{figure*}

Overall, the analysis of the diffusion parameter $D$ indicates that the modifications introduced in this work, including the adoption of updated satellite masses and the inclusion of the new moon S/2025~U1, produce predominantly local effects on the dynamics of Uranus's inner system. For most satellites, the differences among the three scenarios remain small throughout the entire range of integration times considered. The most significant variations are observed for Belinda and Perdita, the only pair involved in a persistent 44:43 mean-motion resonance, highlighting the greater sensitivity of this region to changes in the adopted dynamical model. Increases are also identified for Bianca, Cressida, and Desdemona, although they are less pronounced and do not indicate qualitative changes in their orbital behavior. Taken together, these results suggest that the adopted updates affect the local dynamics of some satellites without substantially altering the global architecture or the long-term stability of Uranus's inner satellite system.

\begin{table}[ht]
\centering
\caption{Numerical values of the maximum eccentricity $e_{\max}$ for the 13 inner satellites of Uranus under the three adopted scenarios.}
\label{tab:e_max}
\small
\scalebox{0.85}{
\begin{tabular}{lccc}
\hline
\hline
Moon & Standard mass & Updated masses & Complete system \\
\hline
Cordelia  & 0.000302 & 0.000302 & 0.000302 \\
Ophelia   & 0.009544 & 0.009544 & 0.009544 \\
Bianca    & 0.002929 & 0.000496 & 0.000354 \\
Cressida  & 0.001650 & 0.001365 & 0.000813 \\
Desdemona & 0.002341 & 0.001678 & 0.001496 \\
Juliet    & 0.001315 & 0.001391 & 0.001351 \\
Portia    & 0.000658 & 0.000715 & 0.000627 \\
Rosalind  & 0.000937 & 0.000939 & 0.000939 \\
Cupid     & 0.000673 & 0.000645 & 0.000642 \\
Belinda   & 0.000903 & 0.000902 & 0.000902 \\
Perdita   & 0.003652 & 0.003665 & 0.003649 \\
Puck      & 0.000101 & 0.000101 & 0.000101 \\
Mab       & 0.003534 & 0.003534 & 0.003534 \\
\hline
\end{tabular}}
\end{table}

The complementary analysis of the maximum eccentricity $e_{max}$ of the satellites reinforces this interpretation by revealing that the dynamical variations induced by the system modifications are spatially localized. In particular, the Bianca group is found to be the most sensitive to the updates, exhibiting a decrease in $e_{\max}$ throughout the region between the orbits of Bianca and Rosalind. This behavior indicates a reduction in orbital instability in this region, consistent with the variation in the diffusion parameter $D$ observed in Sect. \ref{subsec:orbevostab}.

Therefore, the results of this section show that the system is susceptible to updated mass values for Cordelia, Cressida, and Ophelia. The addition of the newly identified moon S/2025~U1 also dynamically affects the other satellites on short timescales.
\section{Analysis of mean-motion resonances}
\label{sec:res}
\begin{figure}
\centering
\begin{minipage}{0.48\linewidth}
  \includegraphics[width=\linewidth]{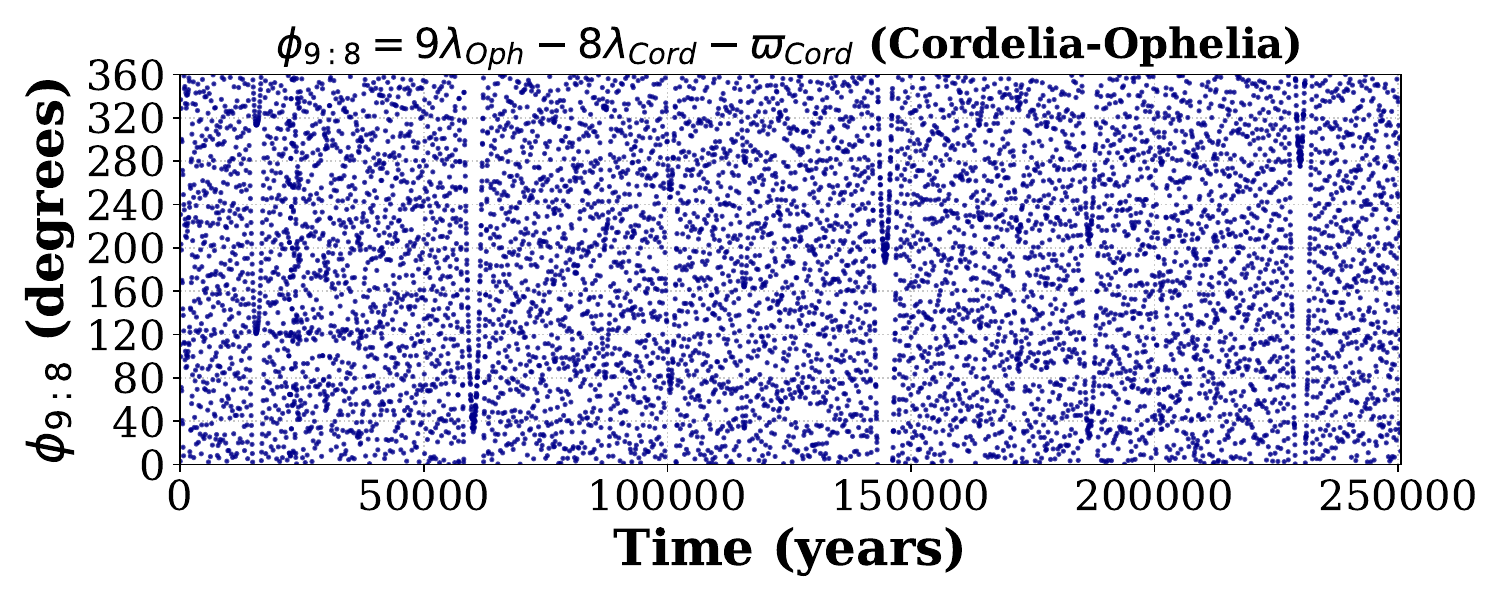}
  \caption*{(a)}
  \end{minipage}
\hfill
\begin{minipage}{0.48\linewidth}
  \includegraphics[width=\linewidth]{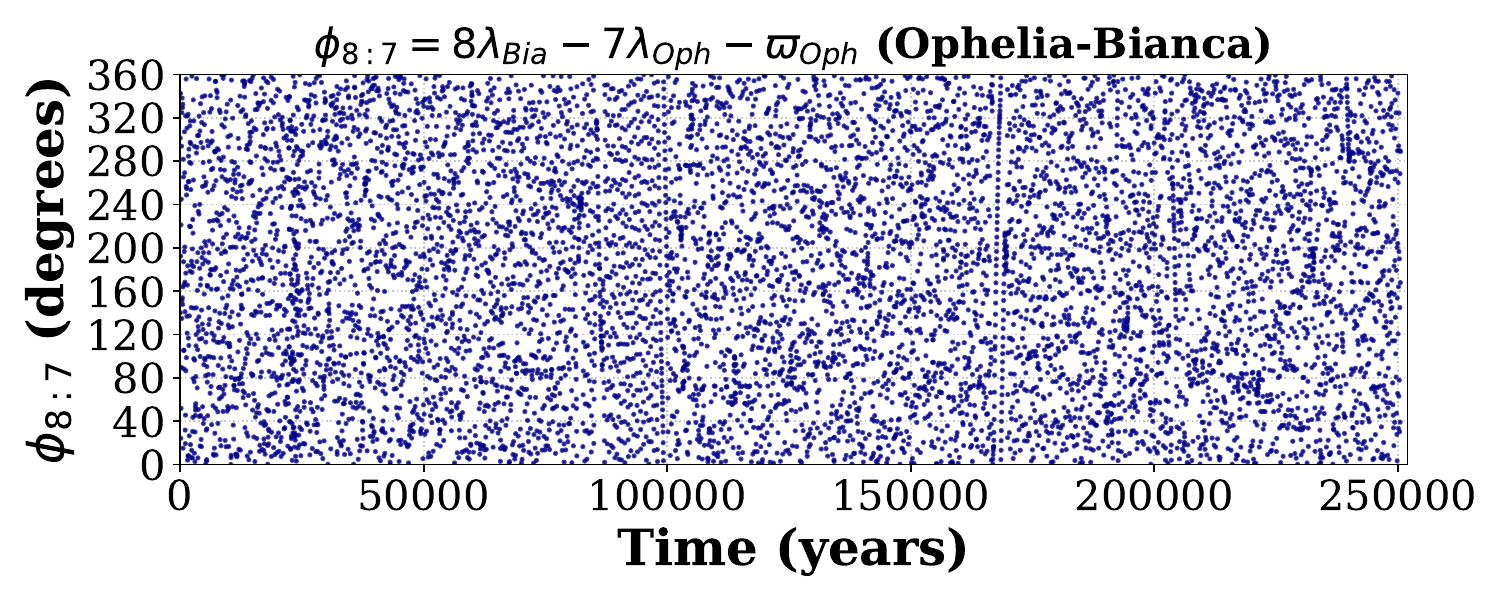}
 \caption*{(b)}
\end{minipage}
\hfill
\begin{minipage}{0.48\linewidth}
  \includegraphics[width=\linewidth]{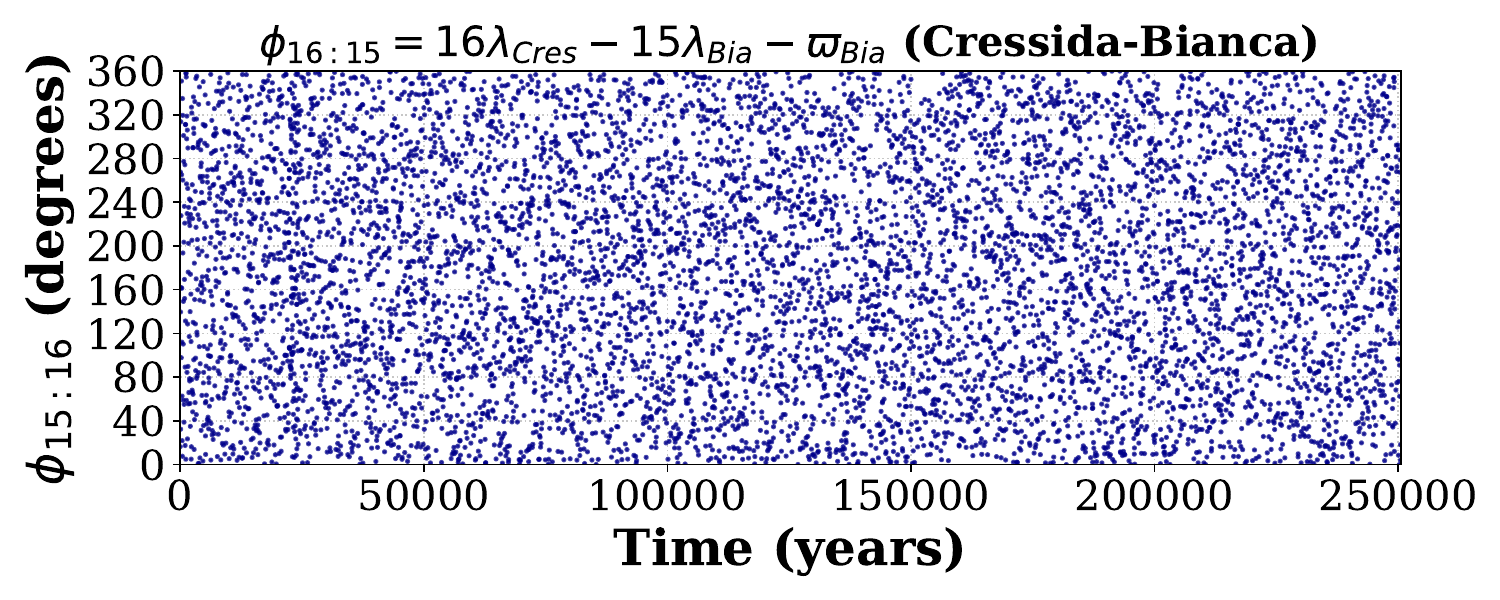}
 \caption*{(c)}
\end{minipage}
\hfill
\begin{minipage}{0.48\linewidth}
  \includegraphics[width=\linewidth]{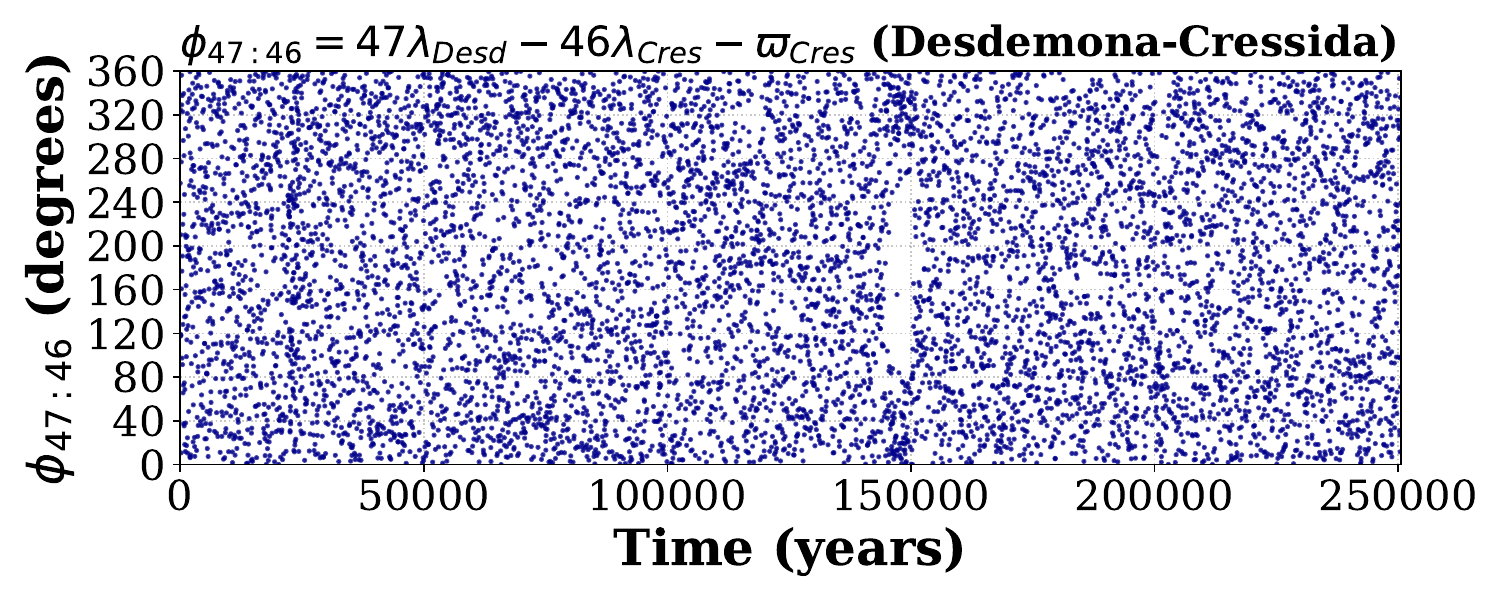}
 \caption*{(d)}
\end{minipage}
\hfill
\begin{minipage}{0.48\linewidth}
  \includegraphics[width=\linewidth]{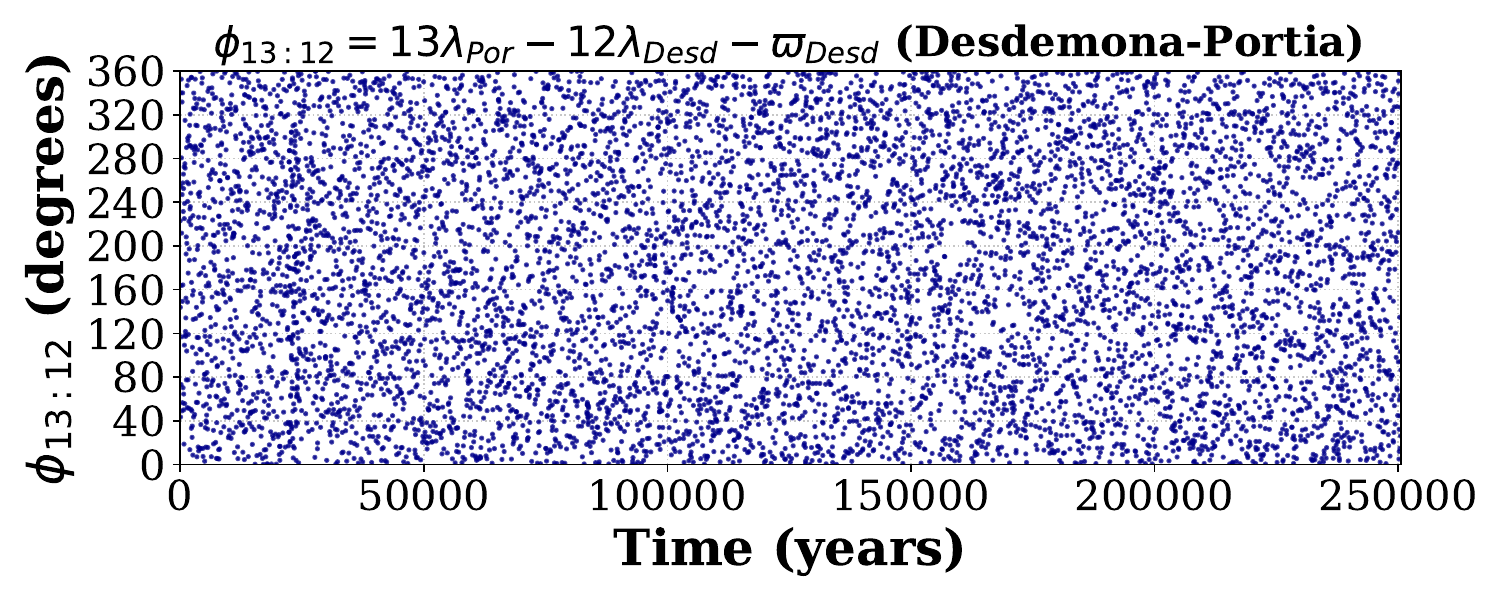}
 \caption*{(e)}
\end{minipage}
\hfill
\begin{minipage}{0.48\linewidth}
  \includegraphics[width=\linewidth]{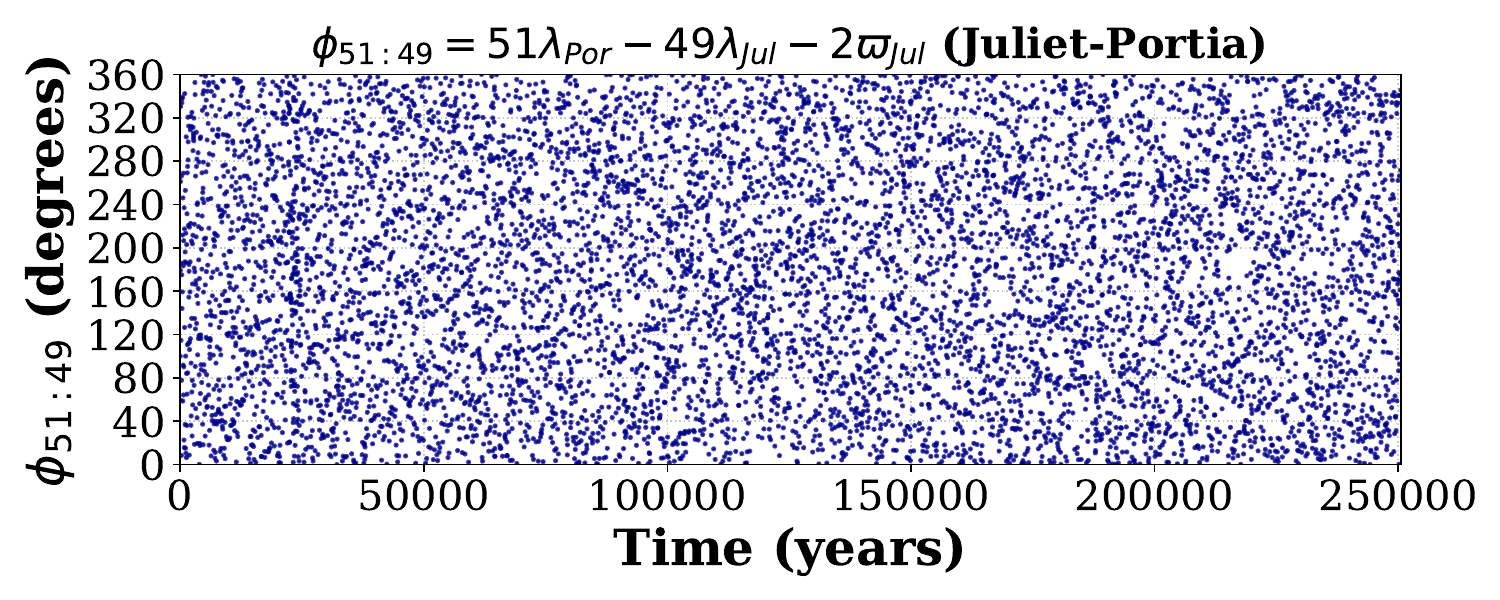}
   \caption*{(f)}
\end{minipage}
\hfill
\begin{minipage}{0.48\linewidth}
  \includegraphics[width=\linewidth]{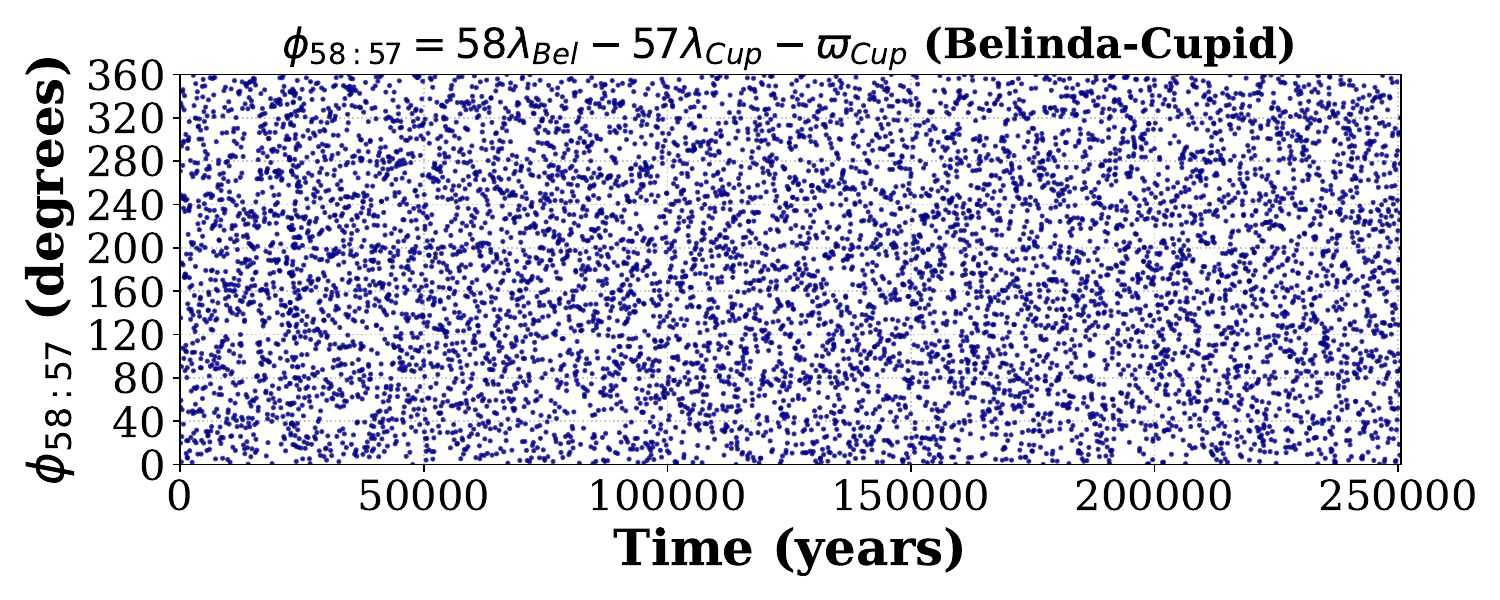}
   \caption*{(g)}
\end{minipage}
\hfill
\begin{minipage}{0.48\linewidth}
  \includegraphics[width=\linewidth]{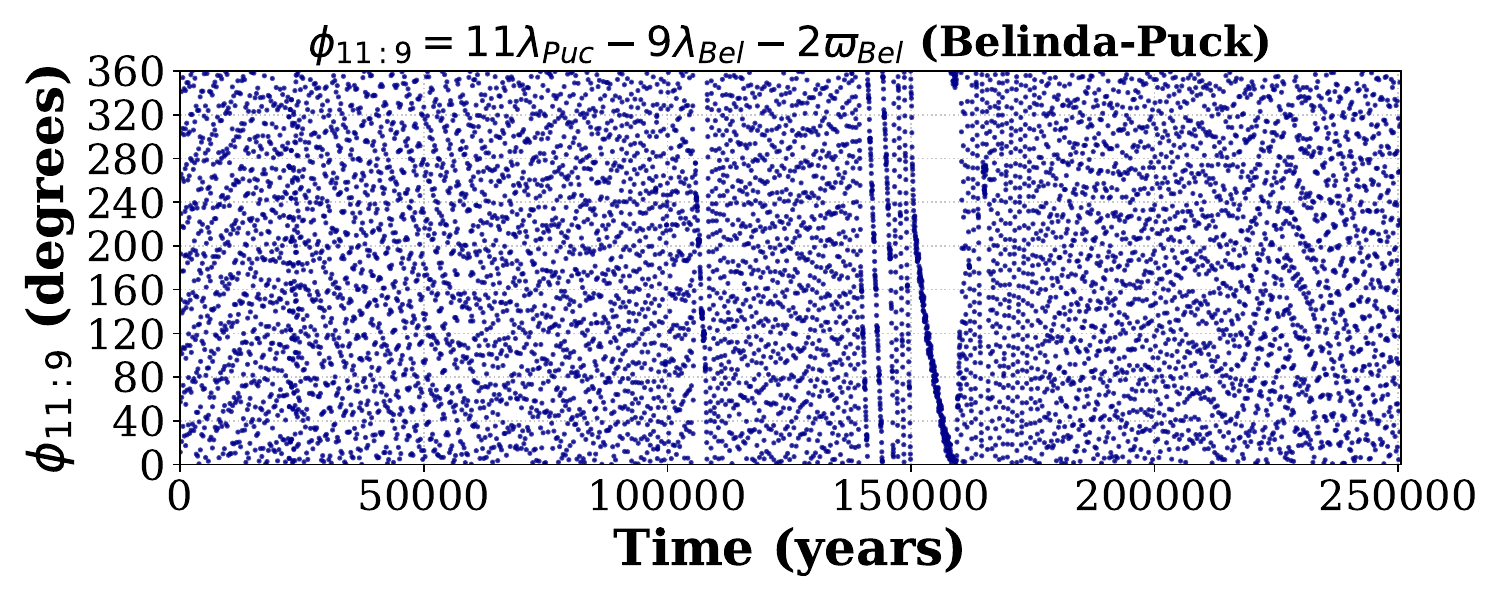}
   \caption*{(h)}
\end{minipage}
\hfill
\begin{minipage}{0.48\linewidth}
  \includegraphics[width=\linewidth]{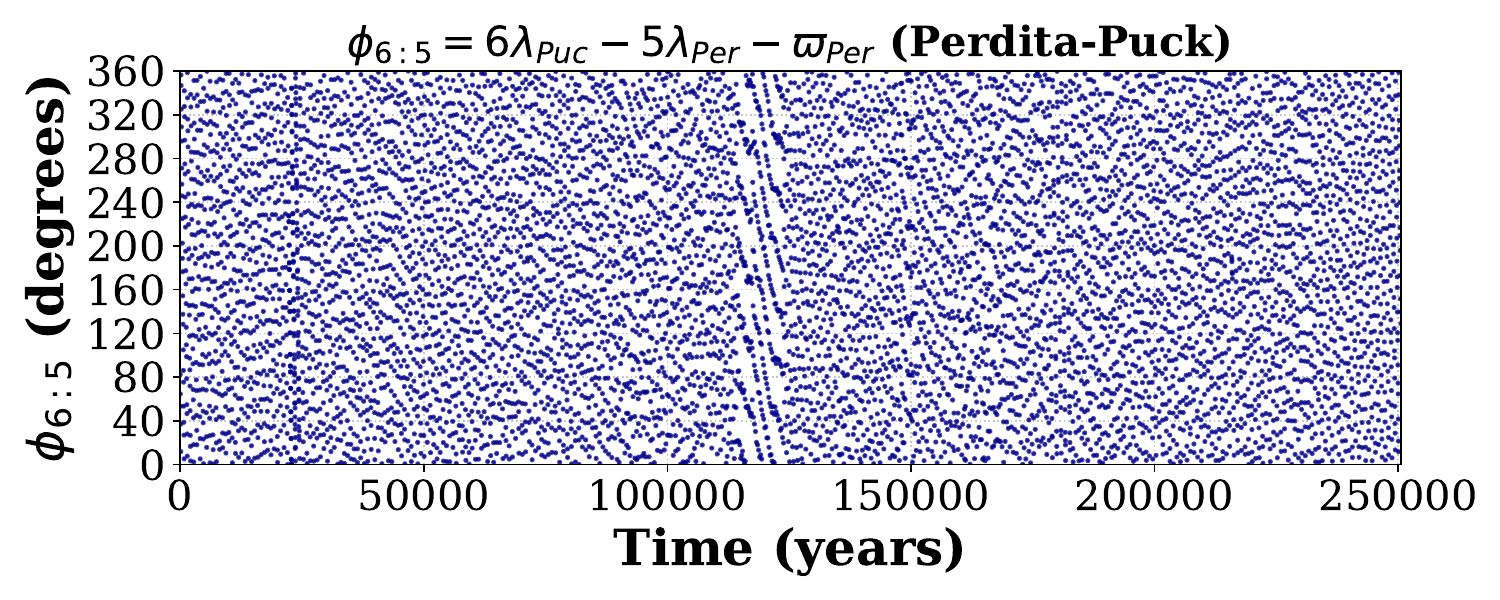}
   \caption*{(i)}
\end{minipage}
\hfill
\begin{minipage}{0.48\linewidth}
  \includegraphics[width=\linewidth]{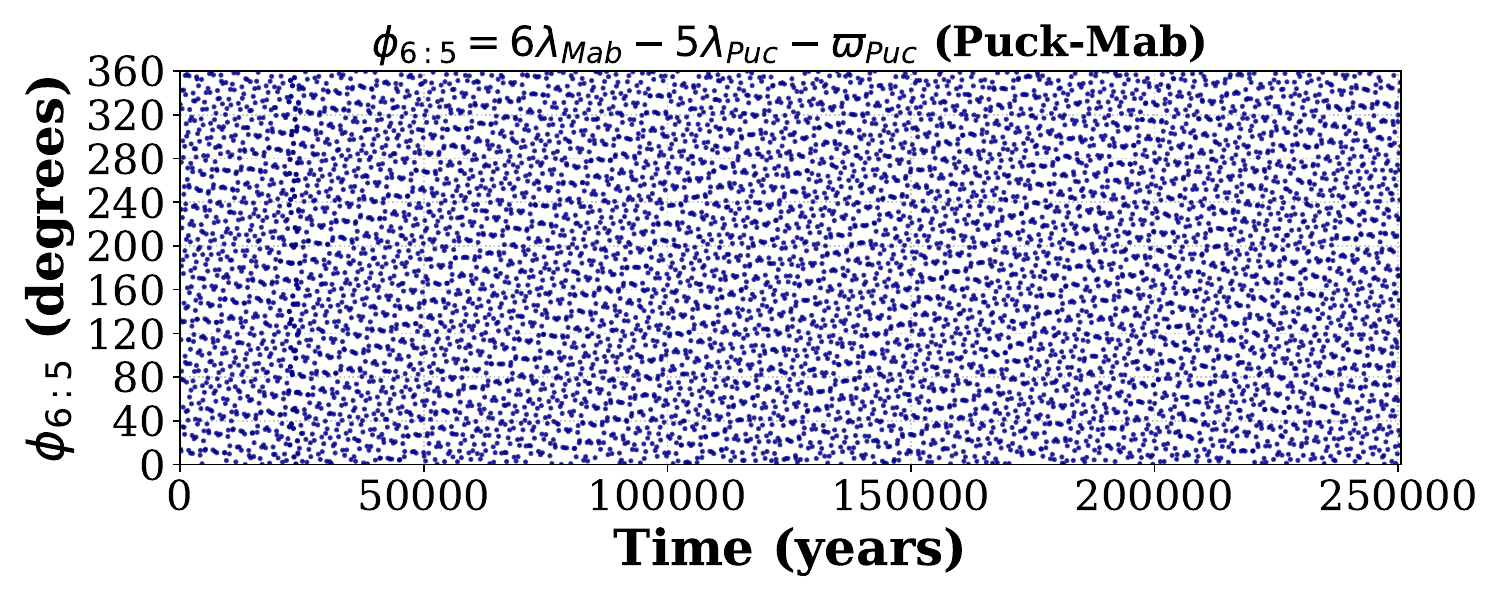}
   \caption*{(j)}
\end{minipage}
    \caption{Temporal evolution of the resonant arguments associated with the main mean-motion resonances identified among Uranus's inner satellites, as indicated in Fig. \ref{fig:All_Ress}. The integrations were performed considering the complete system, including the updated masses of Cordelia, Ophelia, and Cressida, the new moon S/2025~U1, and the updated values of Uranus's $J_2$, $J_4$, and $J_6$.} 
    \label{fig:anal_ress}
\end{figure}

Previous studies have investigated the stability of Uranus's inner satellites using mean-motion resonances \citep{Charalambous2022, French2015, Quillen2014}. The results show that, in many cases, the satellites are not located in the resonant configuration but rather in its vicinity, where significant dynamical interactions can still occur. Furthermore, \citet{Cuk2022} suggests that changes in satellite mass can influence their capture or escape from resonant configurations. Given the updated mass values for Cordelia, Cressida, and Ophelia, updates of Uranus's gravitational coefficients and inclusion of the new moon, we analyzed the main mean-motion resonances (MMRs) involving these bodies. The investigation was restricted to the resonances highlighted in Fig.~\ref{fig:All_Ress}, which was constructed based on the aforementioned studies. We studied the MMRs between two satellites where the resonant argument, $\phi$, generally has the form
\begin{equation}
\phi_{q:(q-1)} = q\,\lambda_j - (q - 1)\,\lambda_i - \varpi_i    
\end{equation} 
and 
\begin{equation}
\phi_{q:(q-2)} = q\,\lambda_j - (q - 2)\,\lambda_i - 2\varpi_i,   
\end{equation} 
\noindent where $\lambda_j$ is the mean longitude of the outer satellite, $\lambda_i$ is the mean longitude of the inner satellite, and $\varpi_i = \omega_i + \Omega_i$ is the longitude of the pericenter of the inner satellite. The integer $q$ defines the resonance ratio (e.g., $q:(q-1)$), but it does not directly correspond to the resonance order.

In this formulation, we analyzed first- and second-order resonances. All analyzed resonance arguments are shown in Fig.~\ref{fig:anal_ress}, and the satellite pairs involved in each resonance are indicated in the figure legends. The integrations were carried out using the complete system, incorporating the updated masses of Cordelia, Ophelia, and Cressida, the newly identified moon S/2025 U1, and the revised values of Uranus's $J_2$, $J_4$, and $J_6$.

Each panel shows the time evolution (in years) of a specific resonant argument, $(\phi)$, constructed from combinations of the
mean longitudes $(\lambda)$ and longitudes of pericenter $(\varpi)$ of the respective satellite pairs. A librating behavior (oscillation around a fixed value) indicates the presence of an actual resonance,
whereas circulation (complete variation from $0^\circ$ to $360^\circ$) suggests the absence of resonance. The resonant angle and the satellite pair analyzed are explicitly identified in the figure labels.

The resonant argument $\phi_{9:8} = 9 \lambda_{Oph} - 8 \lambda_{Cord} - \varpi_{Cord}$, presented in Fig.~\ref{fig:anal_ress}a, is associated with the proximity of a possible 9:8 mean-motion commensurability between Cordelia (Cord) and Ophelia (Oph), with $q = 9$. Although the angle exhibits apparent irregularities or more abrupt variations at specific epochs (around $50,000$, $150,000$, and $250,000$~years), it is evident that it circulates continuously throughout the entire analyzed interval, without showing any libration behavior.

These short-term variations correspond to changes in the circulation induced by neighboring satellites in the densely packed inner Uranian system.
The absence of libration indicates that there is no resonant trapping, and therefore the dynamical interactions associated with this commensurability remain weak.

The resonant arguments analyzed below, shown in panels (b) to (g), also do not exhibit any indication of libration behavior; instead, they circulate continuously from 0$^\circ$ to 360$^\circ$ throughout the entire time interval studied. These arguments are: $\phi_{8:7} = 8 \lambda_{Bia} - 7 \lambda_{Oph} - \varpi_{Oph}$, associated with the possible resonance between Ophelia and Bianca (Bia); $\phi_{16:15} = 16 \lambda_{Cres} - 15 \lambda_{Bia} - \varpi_{Bia}$, between Cressida (Cres) and Bianca; $\phi_{47:46} = 47 \lambda_{Desd} - 46 \lambda_{Cres} - \varpi_{Cres}$, between Desdemona (Desd) and Cressida; $\phi_{13:12} = 13 \lambda_{Por} - 12 \lambda_{Desd} - \varpi_{Desd}$, between Desdemona and Portia (Por); $\phi_{51:49} = 51 \lambda_{Por} - 49 \lambda_{Jul} - 2 \varpi_{Jul}$, between Juliet (Jul) and Portia; and $\phi_{58:57} = 58 \lambda_{Bel} - 57 \lambda_{Cup} - \varpi_{Cup}$, between Belinda (Bel) and Cupid (Cup). The absence of libration in all of these cases indicates that the corresponding mean-motion resonances do not exhibit resonant capture in the complete system configuration considered in this work.

The angle $\phi_{11:9} = 11 \lambda_{Puc} - 9 \lambda_{Bel} - \varpi_{Bel}$, associated with the Belinda-Puck (Puc) pair, Fig.~\ref{fig:anal_ress} (h), also shows a circulating behavior, but exhibits a brief episode of libration in  $\sim100,000$~years, returning to circulation after this interval. This transient libration could be attributed to slow secular variations in the orbital frequencies, induced by weak, cumulative perturbations from other satellites in the inner system. Such variations may temporarily shift the system closer to the commensurability condition, producing a short-lived modulation of the resonant angle. As discussed in Sect.~\ref{subsec:orbevostab}, short-lived resonant episodes of this kind can introduce additional spectral frequency resolutions that affect the estimation of the diffusion parameter $D$ when intermediate integration times are considered, without implying a qualitative change in the long-term dynamical regime of the system.

This behavior may be influenced by the proximity of the mean motion resonance 44:43 between Perdita and Belinda, as well as by the strong perturbations of Puck, the most massive body in the region. A similar evolution of the resonant argument was reported in previous works \citep{French2015,Ricardo2022}.

Analyzing the temporal evolution of the resonant argument $\phi_{6:5} = 6\lambda_{Puc} - 5\lambda_{Per} - \varpi_{Per}$ between Perdita (Per) and Puck (Fig.~\ref{fig:anal_ress}i), a predominantly circulation behavior is observed, with the angle traversing the entire range between $0^\circ$ and $360^\circ$, indicating the absence of resonant trapping. After 100,000 years, different oscillations are shown, but nothing that can be considered libration or proximity to a resonance.

In the case of the resonant argument $\phi_{6:5} = 6 \lambda_{Mab} - 5 \lambda_{Puc} - \varpi_{Puck}$, shown in Fig.~\ref{fig:anal_ress}j, no evidence of libration is observed. Throughout the analyzed time interval, the angle circulates continuously, covering the entire range from $0^\circ$ to $360^\circ$. This circulation indicates the absence of resonant confinement. It suggests that, despite the nominal proximity of the 6:5 period ratio, the interaction between Puck and Mab does not produce a resonant modulation of their orbital frequencies.

From a dynamical point of view, this behavior may indicate that the dominant spectral frequencies associated with Mab's orbit remain well separated, without the introduction of long-period modulations by this commensurability. This result is consistent with the low diffusion values and their weak temporal dependence found for Mab in Sect.~\ref{subsec:orbevostab}, reinforcing the interpretation that its orbit is in a dynamically non-resonant regime.

The analysis of the resonant arguments presented in this section shows that, in most cases, there is no evidence of libration, indicating the absence of mean-motion resonances between the moons at present. Overall, the results confirm that the system is predominantly non-resonant, with possible resonant interactions occurring only sporadically and on short timescales.

\subsection{Mean-motion resonance between Belinda and Perdita}
\label{sec:mmr:BelPer}
Among all the resonant arguments analyzed, only the 44:43 resonance argument between Belinda and Perdita exhibited libration behavior. For this reason, we dedicate this section to a detailed study of the dynamics involving this resonance.

\begin{figure}
\begin{minipage}{0.48\linewidth}
 \includegraphics[width=8.65cm]{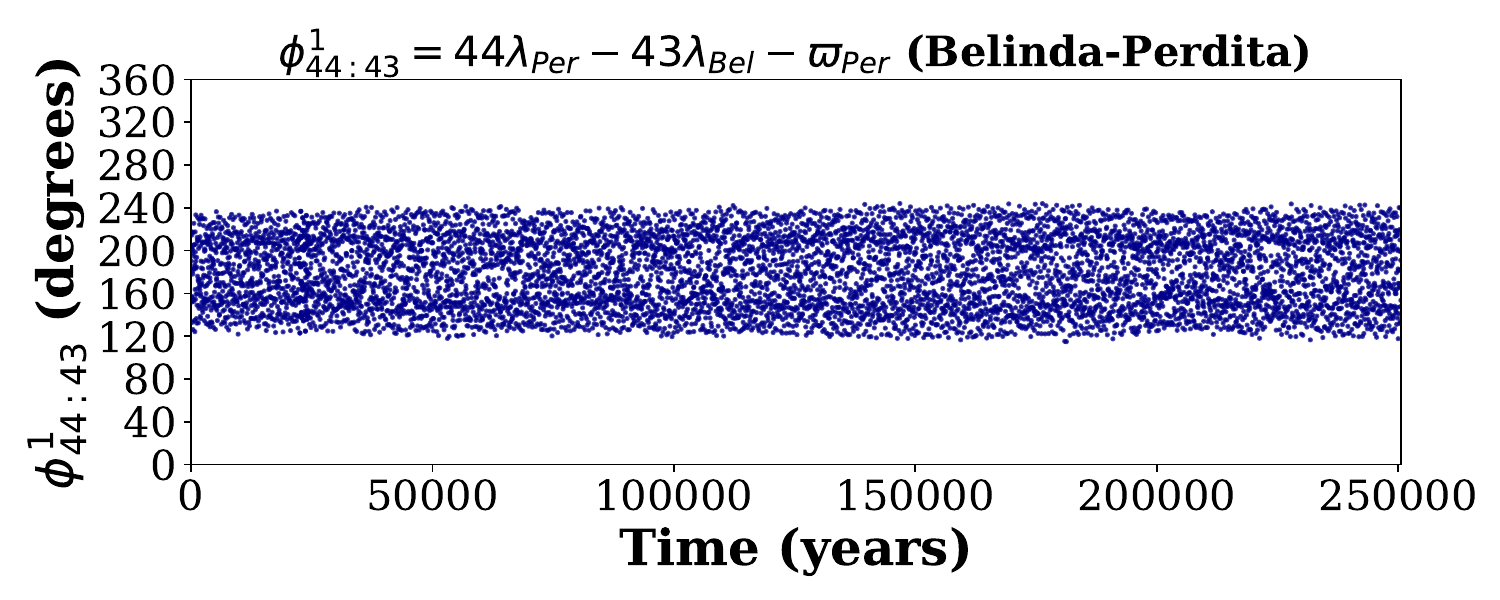}
 \caption*{(a)}
 \end{minipage}
 \\
 \begin{minipage}{0.48\linewidth} 
    \includegraphics[width=8.52cm]{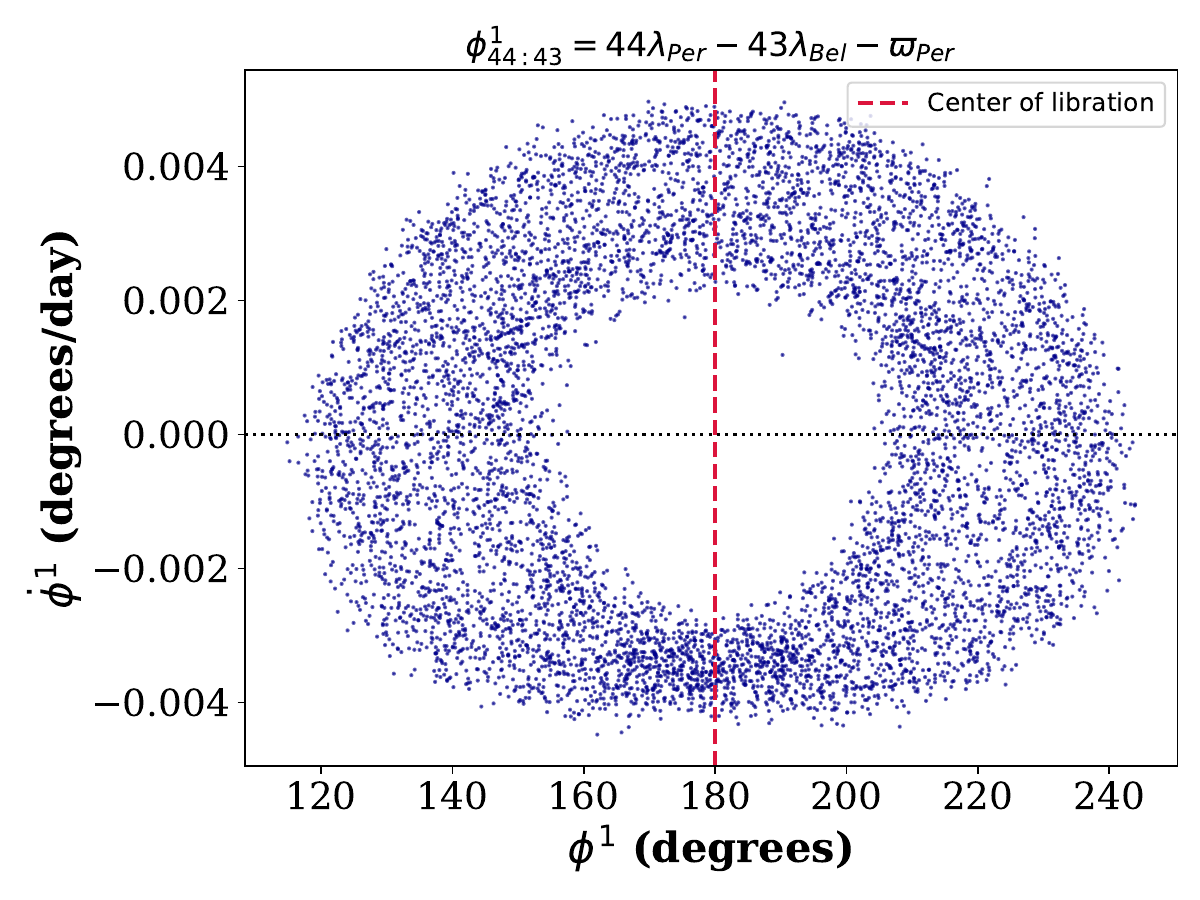}
    \caption*{(b)}
   \end{minipage}
 \caption{Analysis of the mean-motion resonance between Belinda and Perdita. \textit{Panel (a)}: Temporal evolution of the resonant angle $\phi^{1}_{44:43}$. \textit{Panel (b)}: Phase space diagram of the resonant angle $\phi^{1}_{44:43}$ for 250,000~years. The plot shows the evolution of $\phi^{1}_{44:43}$ as a function of its time $\dot{\phi}$, highlighting a libration center near $180^\circ$. The phase-space structure reveals a stable libration pattern, indicating that the system remains in resonance over both short  and long timescales. The subscripts $B$ and $P$ denote Belinda and Perdita, respectively. The integrations were performed including the updated masses of Cordelia, Ophelia, and Cressida; the new moon S/2025~U1; and the updated values of Uranus's $J_2$, $J_4$, and $J_6$.}

\label{fig:fase}
\end{figure}

The analysis of the resonant arguments presented in Fig.~\ref{fig:fase} reveals an interesting dynamical behavior between the satellites Perdita and Belinda. The argument $\phi^{1}_{44:43} = 44\lambda_P - 43\lambda_B - \varpi_P$ shows a libration pattern, oscillating around a fixed value throughout the entire time interval. Although this resonant argument involves the longitude of periapsis of the outer satellite, Perdita, it is one of the standard first-order resonant arguments associated with the 44:43 commensurability. Among the resonant arguments analyzed for this resonance, this is the only one that exhibits sustained libration throughout the integration interval.

This resonance has already been examined in previous studies, which reached different conclusions about the same argument \citep{Cuk2022,Charalambous2022}. In the study by \citet{Charalambous2022}, the authors state that, for the adopted initial conditions and parameters, the resonance does not exhibit libration, with the resonant argument circulating throughout the 35-year interval analyzed. In contrast, \citet{Cuk2022} shows that the argument $\phi^{1}_{44:43}$ librates around a fixed value for up to 50 million years, emphasizing that this resonance may be responsible for maintaining the stability of the Belinda group (Belinda-Cupid-Perdita). Furthermore, they observe that small variations in the satellite's mass can cause the system to leave the resonant regime, highlighting its high sensitivity.

In our analysis, we observed that the libration amplitude remains around $60^\circ$ for all integration times. This value differs from that reported in previous studies for the same resonance \citep{Cuk2022}. This difference may be related to differences between the dynamical model adopted here and those used in previous studies, including updated satellite masses, revised values of Uranus's zonal harmonic coefficients, and the inclusion of S/2025~U1. A dedicated analysis would be required to determine the individual contribution of each of these factors.

To investigate whether the 44:43 resonance between Perdita and Belinda could exhibit additional resonant arguments with libration behavior, we analyzed other combinations of arguments $\phi^{1}_{44:43}$, for example,  $\phi^{2}_{44:43} = 44\lambda_P - 43\lambda_B - \Omega_B$, $\phi^{3}_{44:43} = 44\lambda_P - 43\lambda_B - \Omega_P$ and $\phi^{4}_{44:43} = 44\lambda_P - 43\lambda_B - \varpi_B$. In contrast to what was observed for $\phi^{1}_{44:43}$, all of these angles circulate continuously from 0$^\circ$ to 360$^\circ$ throughout the entire time interval analyzed, indicating the absence of any resonant capture associated with these combinations; for this reason, they are not presented here.
Figure~\ref{fig:fase}b presents the phase diagram of the resonant argument $\phi^{1}_{44:43} = 44\lambda_P - 43\lambda_B - \varpi_P$, associated with the 44:43 mean-motion resonance between Perdita and Belinda, for an integration of 250,000 years. This diagram shows the evolution of the argument as a function of time, $\dot{\phi}$, highlighting the libration pattern and indicating the approximate location of the libration center around 180$^\circ$, marked by the red dashed line.

In this panel, a uniform ring structure is observed, characteristic of a stable libration regime, with an amplitude of approximately $60^\circ$, consistent with what is shown in Fig.~\ref{fig:fase}a. A slight dispersion along the ring is also noticeable, indicating that, over longer integration times, the system experiences additional perturbations, possibly associated with secular effects, indirect interactions with other satellites, or long-term orbital variations. Overall, the comparison between the two diagrams shows the presence of resonant trapping.

The analysis of the critical angles associated with the main mean-motion resonances allows us to assess not only the presence or absence of resonant states, but also their dynamical influence on the orbital stability of the system. In general, the continuous circulation exhibited by most of the analyzed angles indicates the absence of effective resonant confinement, characterizing weak interactions. 
This behavior is consistent with the dynamical regime identified for most of the system and with the low levels of spectral frequency resolution inferred from the diffusion parameter $D$ analysis presented in Sect.~\ref{subsec:orbevostab}.

The episodes of temporary libration observed in some cases reflect transient resonant captures. These captures are likely modulated by long-period secular interactions or by indirect perturbations from neighboring satellites.

The 44:43 resonance between Belinda and Perdita is the only stable resonant configuration identified, exhibiting libration throughout the entire integration interval. This resonance acts as a dynamical confinement mechanism, introducing long secular periods into the system's dynamics. The presence of these long periods provides a natural explanation for the nontrivial behavior of the diffusion parameter $D$ observed for Belinda and Perdita at intermediate integration times, as discussed in Sect.~\ref{subsec:orbevostab}, without indicating any loss of orbital stability.

\subsection{Maximum eccentricity of test particles}
\begin{figure*}
    \centering
    \includegraphics[width=1.0\linewidth, height=7.5cm]{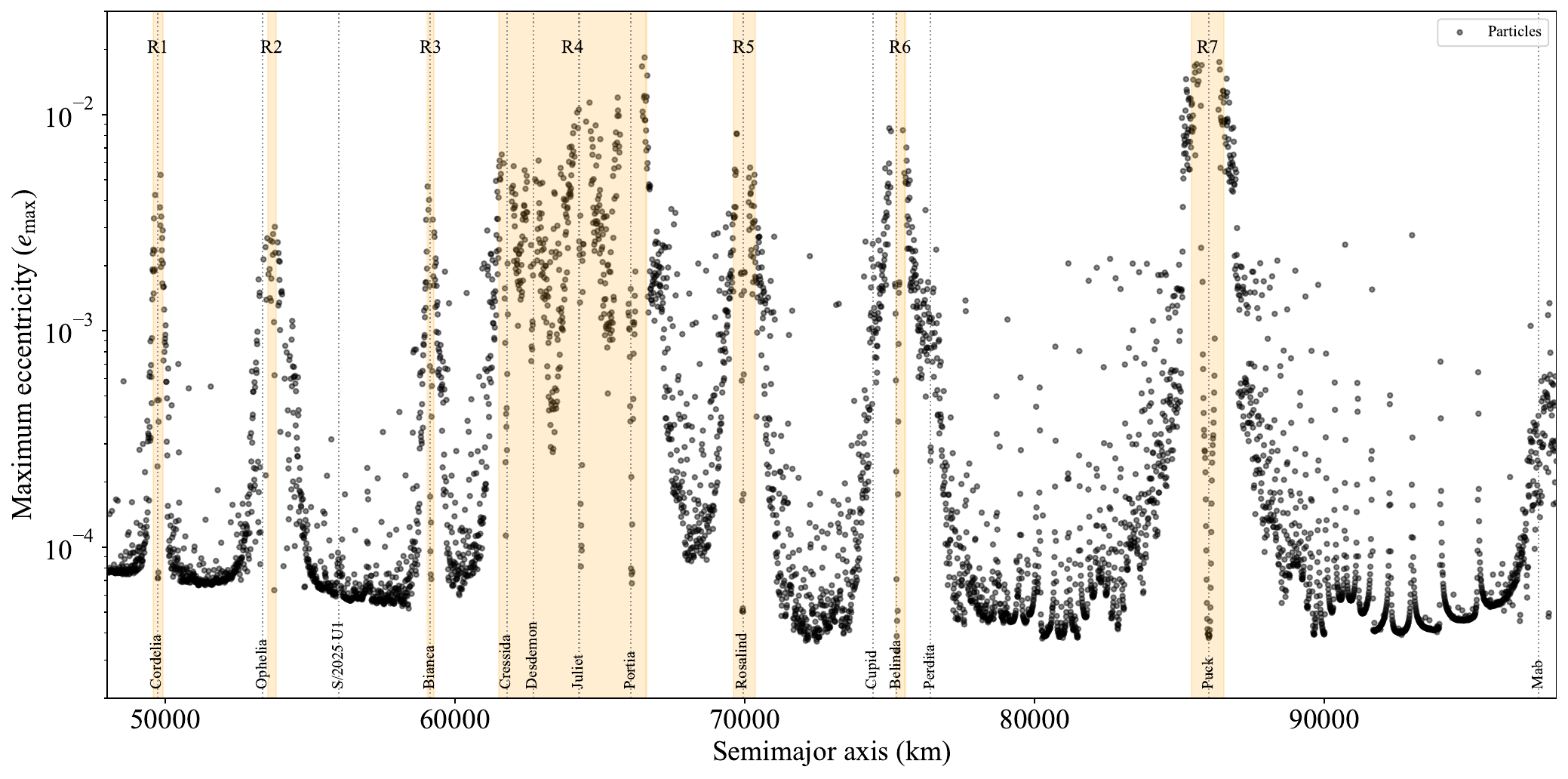}
    \caption{Maximum eccentricity $e_{\max}$ of the test particles as a function of semimajor axis $a$ in the region $48,000 \leq a \leq 98,000$ km with $\Delta a=10$ km. The particles were initially distributed on circular and coplanar orbits, with angular elements $\omega = \Omega = M = 0^\circ$. Gray points represent the maximum eccentricity values. The full system with updated masses and S/2025 U1 is included. Initial conditions of the satellites are from Table~\ref{tab_1}.
The highlighted regions, ranging from R1 to R7, are regions with peaks of maximum eccentricity.}
    \label{fig:e_max_sats}
\end{figure*}
\begin{figure}
    \centering
    \includegraphics[width=1.0\linewidth]{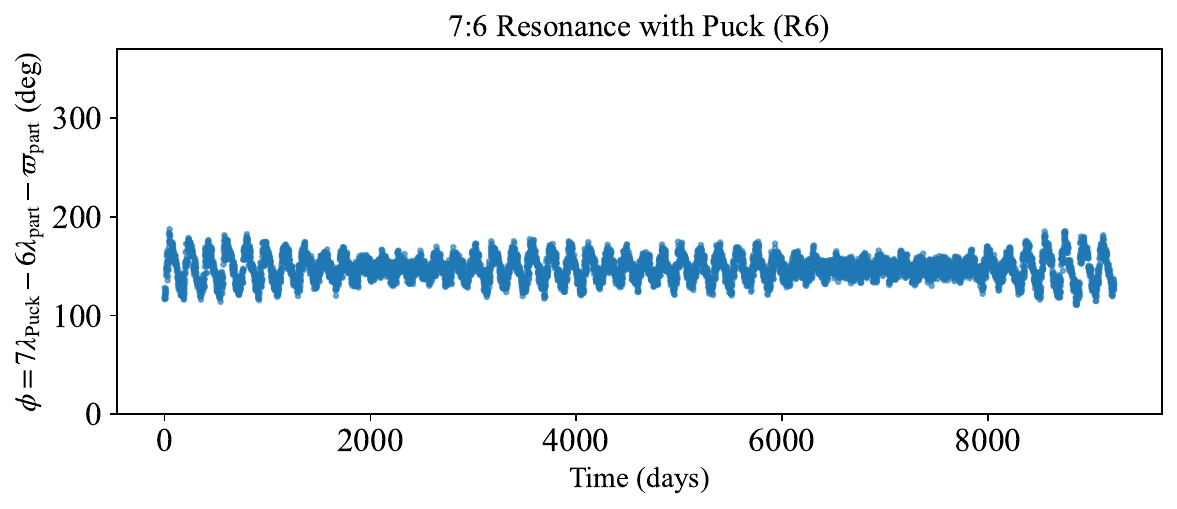}
     \includegraphics[width=1.0\linewidth]{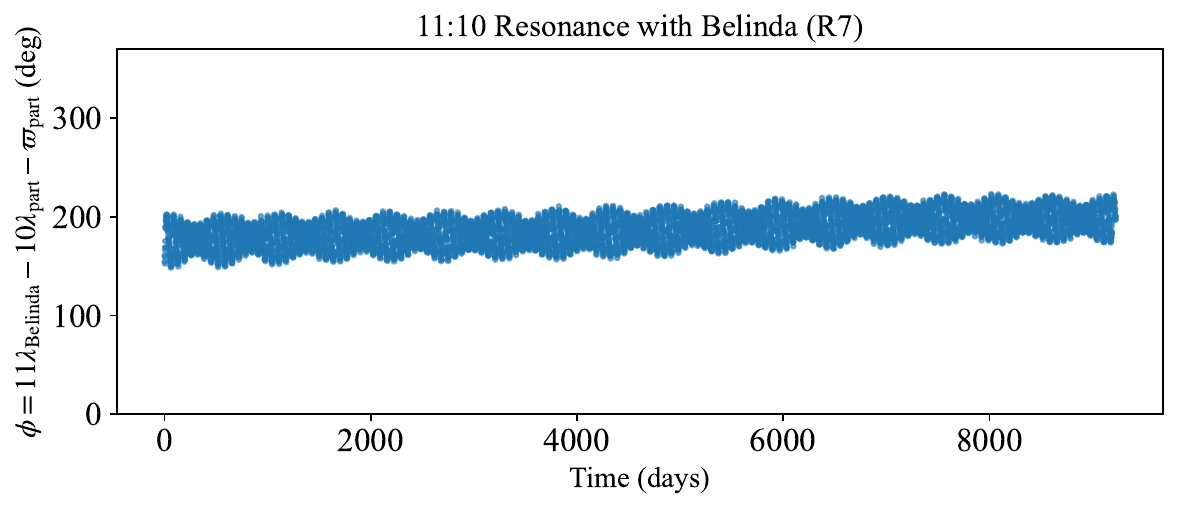}
    \caption{Temporal evolution of the resonant angle for two test particles. First-order resonances 7:6 and 11:10 with Puck and Belinda, respectively, are located in regions R6 (\textit{top}) and R7 (\textit{bottom}). The vertical axis shows the corresponding resonant angle, while the horizontal axis represents time (in days).}
    \label{fig:res_part}
\end{figure}
To investigate changes in orbital stability in the regions between Uranus's inner satellites, we analyzed the maximum eccentricity $e_{max}$ of test particles in relation to these satellites as a function of their semimajor axis $a$. We initially distributed 10,000 particles in circular, coplanar orbits throughout the entire area occupied by Uranus's inner satellites. We then examined the orbital evolution and stability of these test particles by assessing the maximum eccentricity reached over the integration period (10$^4$ Mab periods).

The results, shown in Fig.~\ref{fig:e_max_sats}, reveal a nonuniform structure in the \(a \times e_{\max}\) map, characterized by regions of low and high \(e_{\max}\) values. A relation is observed between increases in \(e_{\max}\) and the proximity to the satellites' orbits, suggesting the action of resonant interactions. This behavior is consistent with the presence of mean-motion resonances previously identified between the particles and the satellites \citep{Charalambous2022}, as illustrated in Fig.~ \ref{fig:All_Ress}.

In particular, distinct regions (R1–R7) can be identified in the vicinity of some satellites, where $e_{max}$ of the particles is significantly enhanced. The region associated with the Portia group (R4) exhibits the most predominant structure, characterized by multiple peaks and overlaps of higher values for $e_{max}$ in a significant radial width in $a$. This behavior reflects both the possible overlap of resonances and the strong gravitational perturbations resulting from the high concentration of satellites within a very compact region.

In contrast, the inner regions (R1–R3) display more isolated peaks and a more regular behavior for $e_{max}$ values. The outer region associated with Puck (R7), however, shows a more defined $e_{max}$ structure, with narrow peaks.

To investigate the origin of these $e_{max}$ structures, we analyzed resonant angles associated with the main resonances identified in regions R1-R7. As representative examples, Fig.~\ref{fig:res_part} shows the resonant angles of two particles corresponding to the first-order resonances 7:6 with Puck and 11:10 with Belinda, located near the highlighted peaks. While regions R6 and R7 exhibit libration, indicating well-established resonant configurations, the resonant angles analyzed in regions R1, R2, R3, R4, and R5 predominantly circulate and that is why they are not presented here. This suggests that, although multiple resonances may be present, the local dynamics in these regions are not governed by a resonant confinement regime. Instead, the observed orbital stability may result from the combined effects of strong gravitational perturbations from the satellites and weak, overlapping, or transient resonant interactions.

In contrast, for particles in regions R6 and R7, first order resonances of the form
$\phi_{q:(q-1)} = q\,\lambda_j - (q-1)\,\lambda_i - \varpi_i$ 
are identified through the temporal evolution of the resonant angles. As an illustration, we present two representative cases in these regions, where libration of the resonant angle is observed, indicating resonant confinement. Most particles analyzed in these regions exhibit qualitatively similar behavior.

The higher $e_{max}$ values occurrence in regions R6 and R7 can be attributed to the combined effect of the 44:43 resonance between Belinda and Perdita and the proximity of Puck. While the resonance between Belinda and Perdita contributes to confinement, the stronger gravitational perturbations induced by Puck likely disrupt the confinement of the particle–satellite mean-motion resonance. This combination appears to favor resonant configurations, in contrast to more compact regions such as R4, where perturbation overlap leads to a more complex $e_{max}$ configuration.

In light of these results, it is further noted that the physical modifications introduced in this work, in particular the updated masses of Cordelia, Cressida, and Ophelia, as well as the inclusion of the new moon S/2025~U1, were not sufficient to establish mean-motion resonances nor to eliminate already existing resonant states. Nevertheless, these changes locally influence the system's dynamical behavior, as observed in the Belinda-Perdita case, through variations in the amplitudes of libration resonant angles (Fig. \ref{fig:fase}). 

It is worth emphasizing that the resonances between satellites analyzed in this work correspond to first- and second-order two-body resonances, whereas, in the case of test particles, only first order two-body resonances were considered. The investigation of other resonant combinations, including different orders and possible multi-body resonances, represents a promising direction for future studies \citep{Quillen2014}.

\section{The dynamical environment of the new moon S/2025~U1}
\label{sec:new_moon}
\begin{figure*}
\begin{minipage}{0.33\linewidth}
  \includegraphics[width=\linewidth]{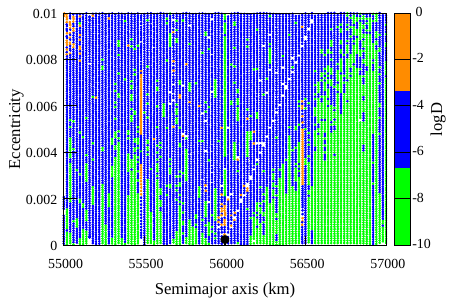}
  \caption*{(a)}
 \end{minipage} 
\hfill
\begin{minipage}{0.33\linewidth}
  \includegraphics[width=\linewidth]{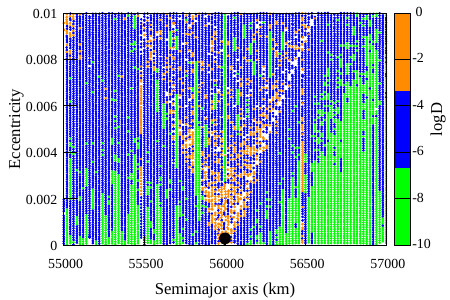}
  \caption*{(b)}
  \end{minipage}
  \hfill
  \begin{minipage}{0.33\linewidth}
  \includegraphics[width=\linewidth]{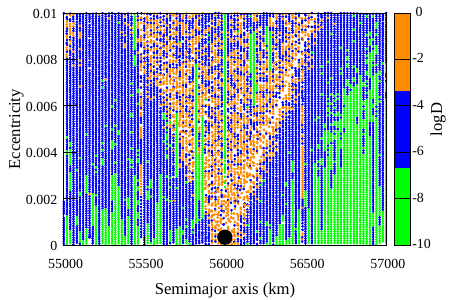}
  \caption*{(c)}
  \end{minipage}
 
\caption{Stability map ($\log D$) of test particles near S/2025~U1 for different assumed radii of the new moon. The particles were distributed according to the following grid: $55,000~\mathrm{km} \leq a \leq 57,000~\mathrm{km}$ with a step of 20~km, $0.0001 \leq e \leq 0.01$ with a step of $0.0001$ and $I=0^\circ$. The angular elements were set to $\omega = \Omega = M = 0^\circ$. The initial conditions for S/2025~U1 were $a = 56,000~\mathrm{km}$, $e = 0.0001$, $I = 0.134^\circ$, and $\omega = \Omega = M= 0^\circ$. \textit{Panel (a)}: S/2025~U1 with $R=3$~km. \textit{Panel (b)}: S/2025~U1 with $R=5$~km. \textit{Panel (c)}: S/2025~U1 with $R=7$~km.}
\label{fig:maps_new_moon}
\end{figure*}
\begin{figure}
    \centering
    \includegraphics[width=1.0\linewidth]{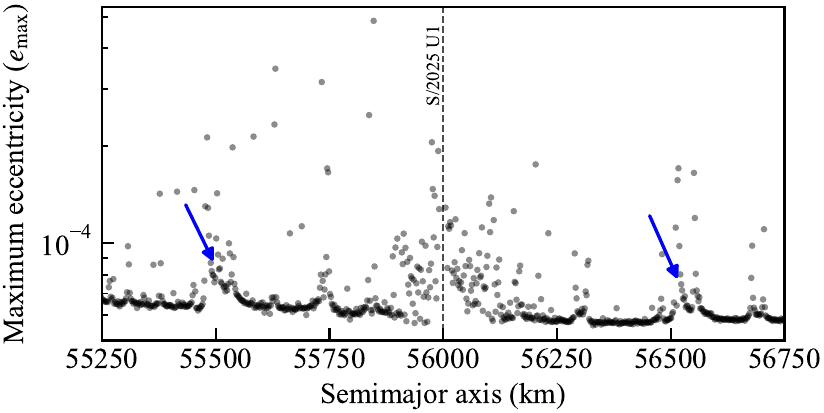}
    \caption{Maximum eccentricity $e_{\max}$ of the test particles as a function of semimajor axis $a$ in the region $55,000 \leq a \leq 57,000$ km. The particles were initially distributed on circular and coplanar orbits with angular elements $\omega = \Omega = M = 0^\circ$. S/2025~U1 was assumed to have a radius of 5 km. Gray points represent individual particles. The blue arrows represent the two regions of high diffusion.
}
    \label{fig:emax_new_moon}
\end{figure}
The peculiar system of Uranus's inner satellites was expanded in August 2025 with the discovery of a new moon, temporarily designated S/2025~U1. Recent observations carried out with the James Webb Space Telescope (JWST) have enabled the detection and preliminary estimation of its physical and orbital parameters, including its approximate radius and mean distance from the planet. S/2025~U1 is located at about 56,000~km from the center of Uranus, between the moons Ophelia and Bianca, and has an estimated radius of roughly 5~km. According to \citet{Maryame2025}, the new moon's orbit is nearly circular, suggesting it likely formed near its current position. These data, provided by the team responsible for the discovery, are summarized in Table~\ref{tab:new_moon}.

\subsection{Nearby dynamics around the new moon}
\label{subsec:nearnewmoon}
In this section we present a detailed analysis of orbital stability near the new moon S/2025~U1 using the diffusion parameter $D$. We used a $a\times e$ grid of particles surrounding S/2025~U1, varying the assumed new moon radius to identify the masses for which the surrounding region remains most stable. We also evaluate, in Sect. \ref{sec:hypsat}, whether such regions could host or favor the detection of additional undiscovered small satellites.

The simulations were performed accounting for the gravitational perturbation of Uranus and all 14 inner satellites of the system (including S/2025~U1), adopting the density proposed by \citet{French2024} ($\rho = 0.9~\mathrm{g/cm^3}$) and radius values in the range $3 \leq R \leq 7$~km. As no observational data on the orbital inclination of S/2025~U1 have yet been reported, we adopted an intermediate value in the simulations based on the mean inclinations of the neighboring moons Ophelia and Bianca to give a more realistic scenario.

The resulting diffusion maps are shown in Fig.~\ref{fig:maps_new_moon}. The numerical simulations were performed over $10^4$ orbital periods of Mab \citep{Laskar2001}. Regions with low diffusion $D$ (green) indicate greater stability of particles' orbits over long periods. In contrast, regions with high diffusion $D$ (orange) indicate lower particle stability during the same amount of time. Areas with intermediate diffusion (blue) reflect moderate orbital stability. The white points represent particles that have collided with the new moon. 

The maps reveal a dynamical behavior that is strongly dependent on the assumed radius of S/2025~U1. For $R = 3$~km, Fig.~\ref{fig:maps_new_moon}a, only very weak orbital instability is observed, with most particles remaining in low diffusion regions and only a few cases of collisions. 

For $R = 5$~km, shown in Fig. \ref{fig:maps_new_moon}b, the dynamical environment remains favorable to orbital stability during the time integration. A broad set of blue points extends on both the inner (to the left of the new moon's orbit) and the outer (to the right of the new moon's orbit) sides, covering nearly the entire range of eccentricities explored. Even in the immediate vicinity of the new moon, blue points appear at very low eccentricities, interspersed with green areas, indicating that near coorbital trajectories can remain regular over long integration times. As the eccentricity increases, this green band progressively narrows and gives way to shades of blue, reflecting a sensitivity to gravitational perturbations, while still allowing the long-term stability of many orbits.
\begin{figure*}[!h]
    \centering
    \begin{minipage}{0.48\textwidth}
        \centering
        \includegraphics[width=\linewidth]{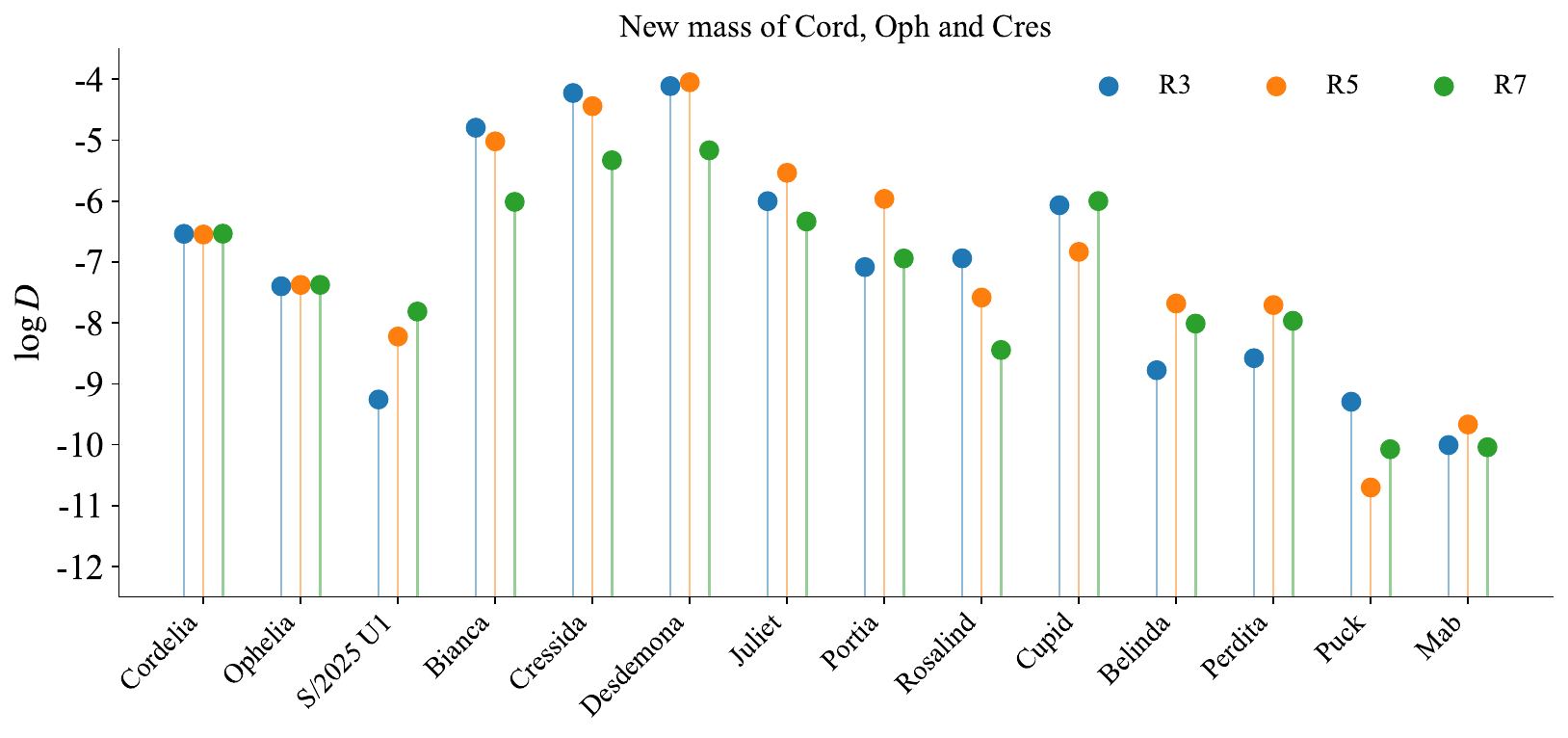}
        \caption*{(a) }
    \end{minipage}
    \hfill
    \begin{minipage}{0.48\textwidth}
        \centering
        \includegraphics[width=\linewidth]{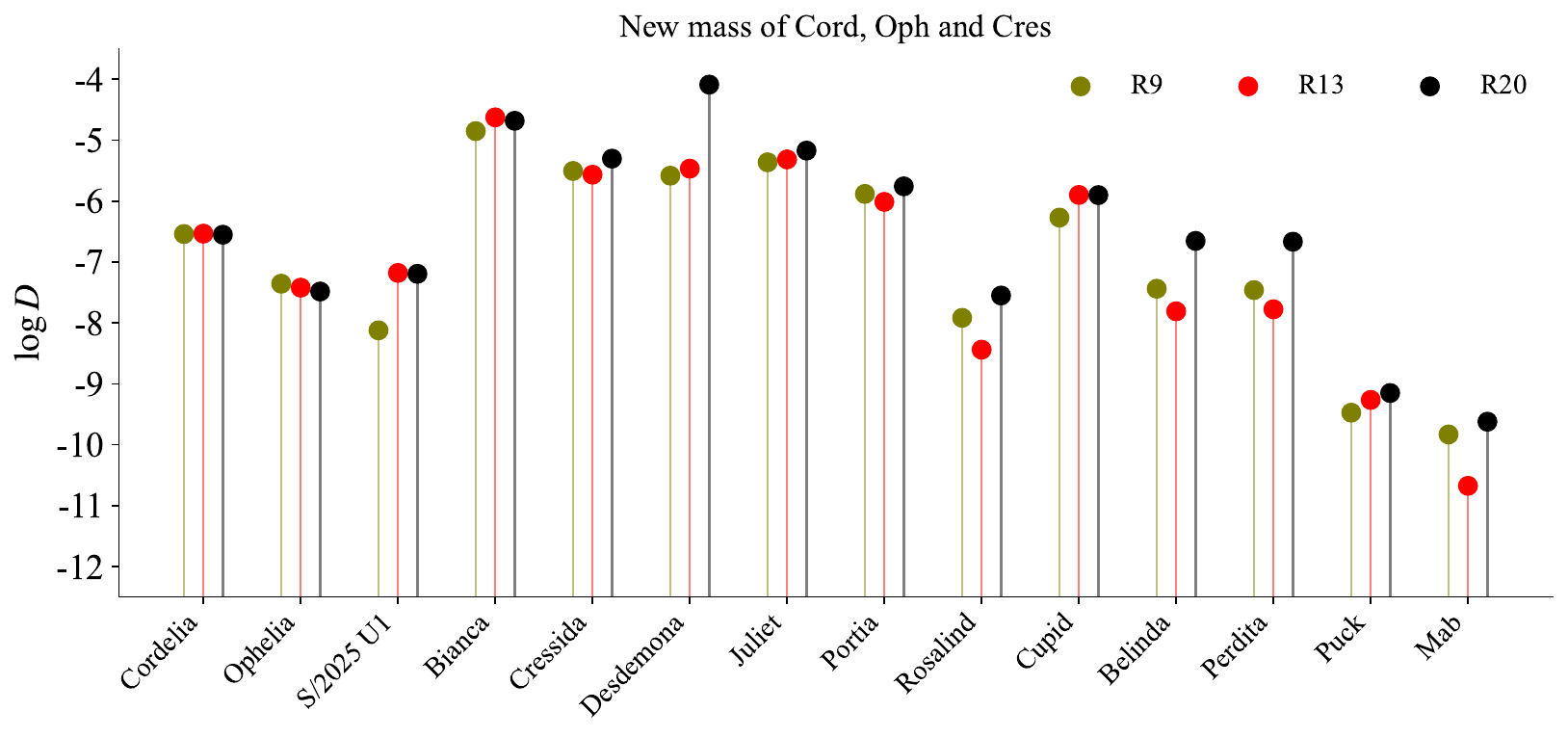}
        \caption*{(b)}
    \end{minipage}
    \caption{Diffusion parameter $\log D$ as a function of various radius values assumed for S/2025~U1 while also considering the updated masses of Cordelia, Ophelia, and Cressida. \textit{Panel (a)}: Radius, $R$, ranging from 3 to 7 km. \textit{Panel (b)}: Radius, $R$, ranging from 9 to 20 km. Each point represents the maximum value of log $D$ for a given satellite. Initial conditions were obtained from Table \ref{tab_3}, adopting the updated masses of Cordelia, Ophelia, and Cressida. Integrations were performed over $10^6$ orbital periods of Mab. }
    \label{fig:global_diffusion}
\end{figure*}

When the radius increases up to $R = 7$~km as shown in Fig. \ref{fig:maps_new_moon}c, the presence of blue points near the new moon is drastically diminished, especially for eccentricities ranging from 0.0 to $\sim0.002$. Although broad stable zones still appear in the inner and outer parts of the map, they shift farther away from the new moon's orbit. At the same time, the intermediate, the region inside and outside the new moon's orbit, still shows regions of stability with $logD$ values between -10 and -4 (blue and green). In this configuration, the new moon's immediate surroundings display only moderate orbital stability; there is no rapid diffusion, but the dynamics are already perturbed enough to prevent the long-term survival of dense or particle populations. Despite the gradual increase in unstable regions near the satellite with R = 7 km, the inner and outer regions of low diffusion persist. In Sect. \ref{sec:hypsat} we investigate whether these regions are capable of hosting new satellites without compromising the orbital stability of this and other neighboring moons.

The diffusion maps reveal two vertical regions of enhanced diffusion in all analyzed cases, located at $a \simeq 55,500$ km and $a \simeq 56,500$ km. No confinement of mean-motion resonances with known satellites was identified at these locations using FMA, which motivated a more detailed investigation of the dynamical behavior in these regions, particularly in the vicinity of the new moon.

To this end, we zoomed into the corresponding region in Fig. \ref{fig:e_max_sats}, considering the interval $55,000 \leq a \leq 57,000$ km (Fig.~\ref{fig:emax_new_moon}), and analyzed the maximum eccentricity $e_{max}$ reached by each particle over a total integration time equivalent to $10^4$ orbital periods of Mab.

We find that, in the two regions associated with the $e_{max}$ peaks shown in Fig.~\ref{fig:emax_new_moon}, particles cluster near high-order mean-motion resonances with the satellites Ophelia and Bianca. In the inner $e_{max}$ peak, at $a\sim55,500$ km, we identify weak resonances such as $87\!:\!79$ with Bianca, $83\!:\!87$ with Ophelia, and $76\!:\!69$ with Bianca. In the outer $e_{max}$ peak, beyond the orbit of new moon S/2025 U1, the resonances are also in high order and involve Ophelia, including $90\!:\!97$, $103\!:\!111$, and $77\!:\!83$.
\begin{figure*}[!h]
    \centering
    \begin{minipage}{0.48\textwidth}
        \centering
        \includegraphics[width=\linewidth]{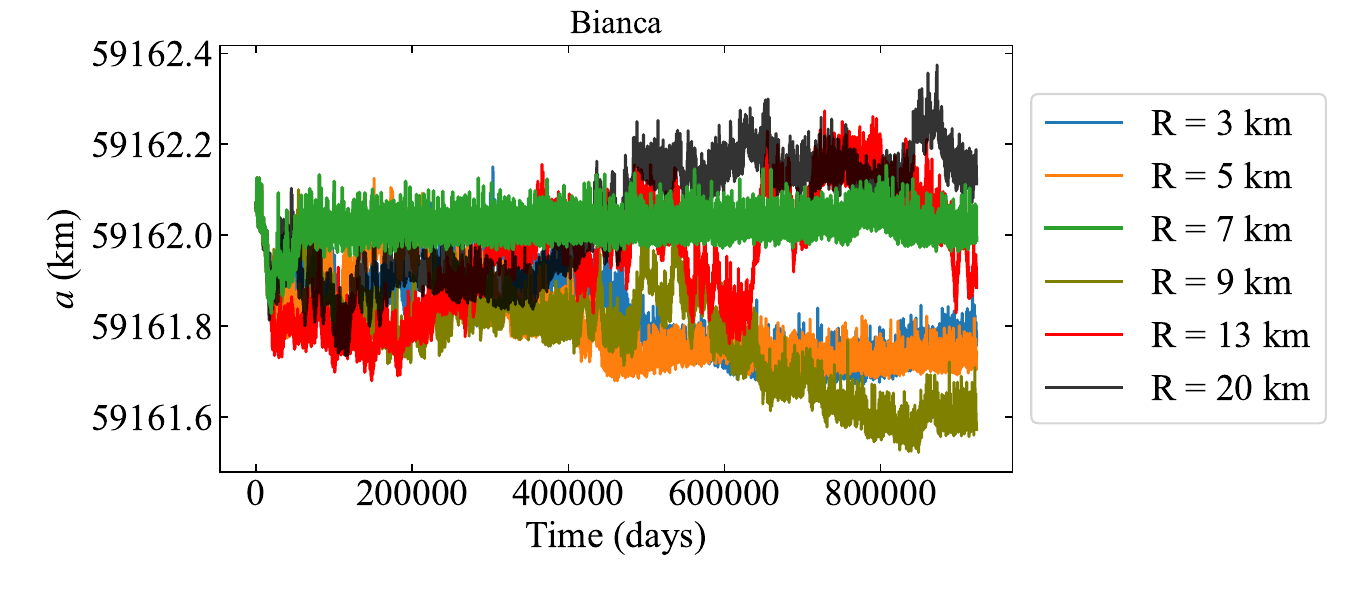}
    \end{minipage}
    \hfill
    \begin{minipage}{0.48\textwidth}
        \centering
        \includegraphics[width=\linewidth]{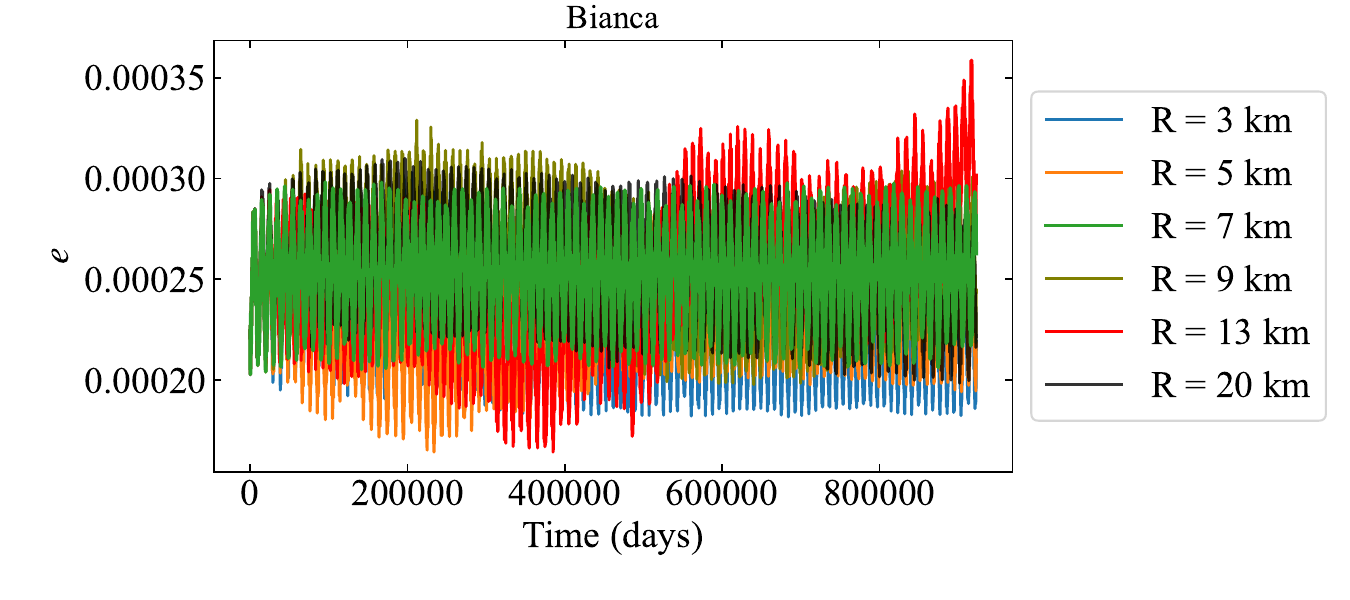}
        \end{minipage}
      \hfill
    \begin{minipage}{0.48\textwidth}
        \centering
        \includegraphics[width=\linewidth]{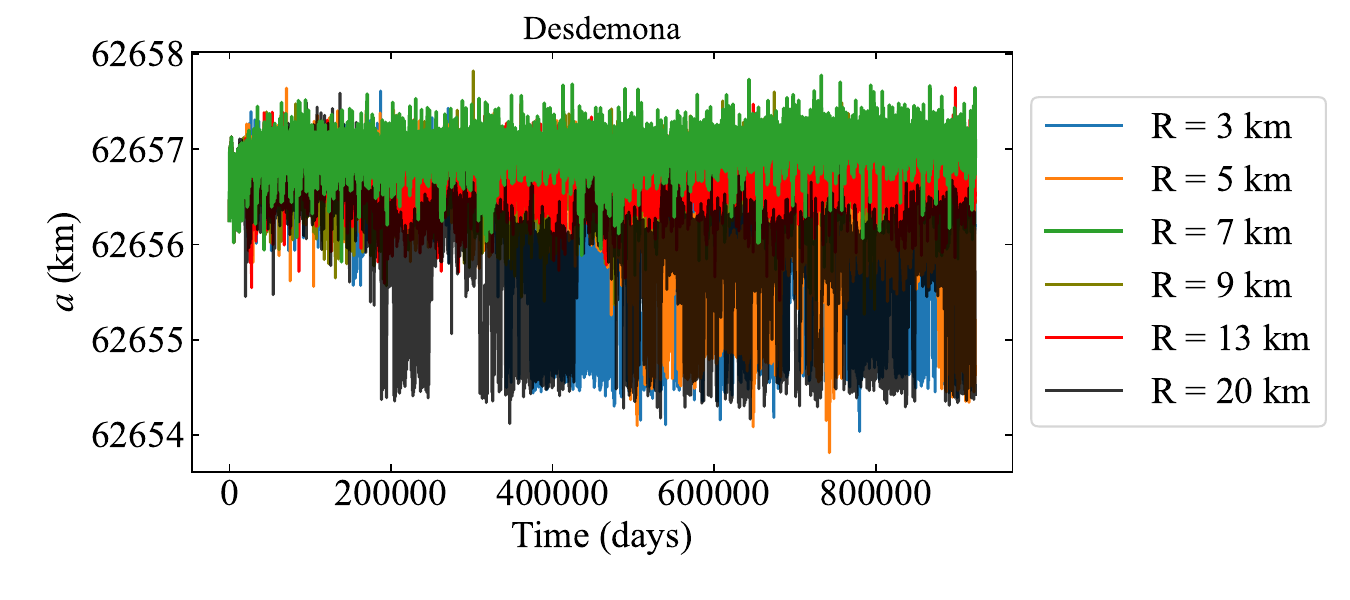}
        \end{minipage}   
 \hfill
    \begin{minipage}{0.48\textwidth}
        \centering
        \includegraphics[width=\linewidth]{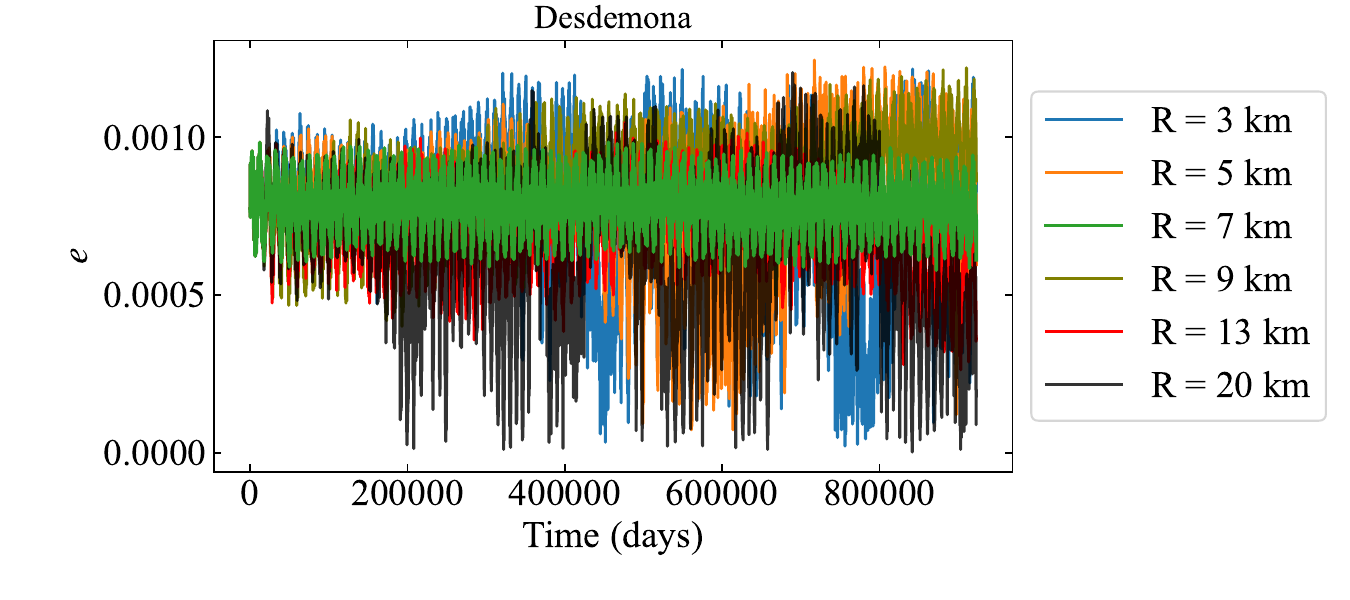}
        \end{minipage}
    \caption{Temporal evolution of the semimajor axis and eccentricity of Bianca (upper panels) and Desdemona (lower panels) for different radius values assumed for the new moon S/2025~U1 ($3 \leq R \leq 20$ km). Initial conditions were obtained from Table \ref{tab_3}, and we adopted the updated masses of Cordelia, Ophelia, and Cressida. Integrations were performed over $10^6$ orbital periods of Mab.}
    \label{fig:ae_bia_desd}
\end{figure*}

Due to the highly compact configuration of the inner Uranian satellite system, these regions are likely permeated by a chain of resonances that may contribute to shaping the observed diffusion structures. As the eccentricity $e_{max}$ increases, possible higher-order two-body resonances and three-body resonances become dynamically relevant, driving a gradual transition from a low-diffusion regime (green) to a high-diffusion regime (orange), as seen in the maps in Fig. \ref{fig:maps_new_moon}.

Across all cases examined within the range $3 \leq R \leq 7$~km, a recurring dynamical pattern emerges. For eccentricities slightly above $e\sim0.002$, and at $a\sim56,000$,km, the same semimajor axis as the newly identified moon, the particles exhibit low $D$ values, regardless of the radius assumed for S/2025 U1. This behavior indicates that these orbits remain stable even under gravitational perturbations, making them candidates for possible coorbital satellites in the vicinity of the new moon. 

In summary, the dynamic environment around S/2025~U1 depends on the assumed satellite radius. While radii smaller than 3–5 km are associated with largely stable configurations, values close to 7 km already lead to partial instability of particles near the moon. Even so, stable regions persist both inside and outside its orbit, indicating that additional particles or small satellites may still be supported.

\subsection{Global stability effects from the new moon}
\label{sec:global_difusion}
In Sect. \ref{subsec:nearnewmoon} we analyze how an increase in the radius ($\equiv$mass) of the newly discovered moon S/2025~U1 may affect the dynamics of test particles distributed in the inner and outer regions of its orbit. However, it is also necessary to investigate how these variations influence the orbits of the other satellites, since they are massive bodies.

In this section we investigate how variations in the radius ($\equiv$mass) of the new moon affect the dynamics of the system's satellites, as well as those of its closest neighbors, Ophelia and Bianca, through the diffusion parameter $D$. For this, we consider a range of radii $3 \leq R \leq 20$ km, corresponding to the values $R = 3, 5, 7, 9, 13$ and $20$ km.

The choice of this band is motivated by the dimensions of some of the system's satellites. After its discovery, S/2025~U1 became the smallest known satellite in this group, with an estimated radius of approximately 5 km. At the same time, previously this position was held by Cupid, whose radius is approximately 9 km. In turn, the satellites neighboring S/2025~U1, Cordelia and Bianca, have radii of approximately 21 km and 27 km, respectively. In addition, we included the intermediate value of 13 km, comparable to the radii of Perdita and Mab. Thus, the adopted band encompasses values smaller than the estimated radius of S/2025~U1, as well as representative sizes observed among the inner satellites of Uranus and those of its neighboring moons. Additionally, the inclusion of radii larger than those considered in the previous section allows us to investigate the effect of higher masses on the orbital stability of the system.

Assuming constant density, we increase the radius of the new moon S/2025~U1, which also implies an increase in its mass values. In this context, we analyze the case in which the radius of S/2025~U1 is varied, using the updated masses of Cordelia, Ophelia, and Cressida (New Mass). The simulations were integrated over $10^6$ orbital periods of Mab.

The results of this analysis are presented in Fig.~\ref{fig:global_diffusion}. For cases with $R$ varying from 3 to 7 km (Fig.~\ref{fig:global_diffusion}a), the highest diffusion values are concentrated in the region occupied by Bianca and nearby satellites. In general, the variations in $\log D$ between the different radius values are small.

However, some satellites exhibit interesting behavior: at $R = 7$ km, a reduction in diffusion is observed, suggesting that a new moon with this radius and mass, in this orbit, may locally stabilize the orbital evolution of the system. This effect is particularly visible for the satellites Bianca, Cressida, Desdemona, Juliet, Portia, and Rosalind.

Analyzing the temporal evolution of the semimajor axis and eccentricity of two neighboring satellites of interest, Bianca and Desdemona (Fig. ~\ref{fig:ae_bia_desd}), the effect associated with the case $R = 7$ km is observed. For both Bianca (upper panels) and Desdemona (lower panels), the variations in the semimajor axis and eccentricity remain approximately constant when compared to other radius values.

This behavior indicates that under the influence of a moon with a radius of 7 km orbiting S/2025~U1, the orbits of these satellites exhibit more regular evolution. This suggests a possible local stabilization of the system.

For the other satellites, it is observed that for larger values of $R$ (9, 13, and 20 km), the diffusion increases slightly. However, most satellites show similar diffusion values across all analyzed radii.

One notable case is Desdemona, which exhibits a significant increase in diffusion at radius $R = 20$ km. In contrast, its diffusion decreases at $R = 9$ and $R = 13$ km.

This behavior suggests that a moon with a radius of 20 km in the orbit of S/2025~U1 may induce more intense perturbations on satellites in the outer orbit, such as Desdemona, affecting their orbital stability. On the other hand, at radii $R = 9$ and $R = 13$ km, a more regular behavior is observed, consistent with possible orbit stabilization, similar to the case at $R = 7$ km.

For the moons neighboring the inner orbit of the new moon, the variations in the diffusion parameter are negligible. Cordelia and Ophelia show little variation, even at $R = 13$ km and $R = 20$ km.

For the outermost satellites, starting from Juliet, the increase in radius from S/2025~U1 does not produce relevant changes. Puck and Mab show some variations in log $D$, but still with very low diffusion values indicating orbital stability.

The reduced diffusion observed for $R = 7$ km may be associated with a more regular dynamical regime, possibly linked to resonant interactions that confine the orbital evolution of nearby satellites. Overall, the results indicate that a satellite of this size orbiting S/2025~U1 is associated with a more regular evolution of the neighboring satellites in the Bianca group, suggesting a possible local stabilization of the group.

For the nearby satellite Ophelia, only a slight variation in diffusion is observed for $R \geq 9$ km. However, at $R = 20$ km, the neighboring satellites of S/2025~U1 begin to show a more significant increase in the diffusion parameter $D$, indicating the onset of a more dynamically perturbed orbital regime.

\subsection{Orbital stability and evolution about the new moon}
\label{subsec:dynnewmoon}
\begin{figure}
    \centering
    \includegraphics[width=1.0\linewidth]{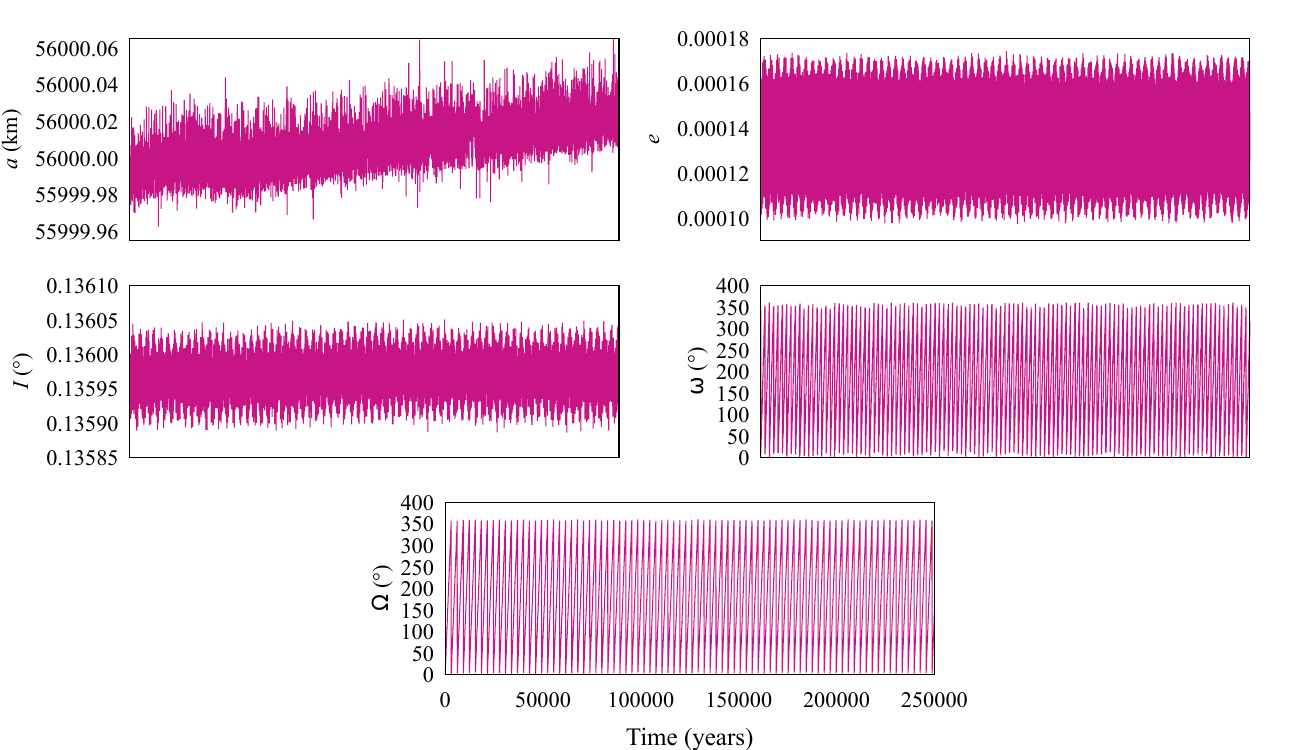}
    \caption{Temporal evolution of the six orbital elements of S/2025~U1 over 250,000~years, computed assuming a radius of 5~km, the value that best matches the dynamical constraints obtained in this study. $a_i=56,000$~km, $e_i=0.0001$, $I = 0.134^\circ$, $\omega = \Omega = M= 0^\circ$.}
    \label{fig:aei_U1}
\end{figure}

We also analyzed the orbital stability and evolution of the new moon S/2025~U1 to understand how its inclusion in the system affects Uranus's inner satellite system.
Figure \ref{fig:aei_U1} shows the evolution of the orbital elements of S/2025~U1 over approximately 250,000~years, assuming a radius of 5~km for the new moon. This value was adopted because it is consistent with the one used by the team that discovered the new moon.

The evolution of semimajor axis $a$ remains confined around 56,000~km, with only minor variations throughout the simulated interval. This orbital stability indicates that the new moon does not undergo significant orbital evolution and that mutual perturbations with Ophelia and Bianca are limited.

The evolution of eccentricity $e$ remains very low (on the order of $10^{-4}$) and exhibits periodic oscillations of small amplitude. This suggests that the orbit of S/2025~U1 remains nearly circular, consistent with the behavior of the other inner satellites. The absence of any significant increase in eccentricity suggests that the orbit of the new moon remains dynamically stable for the initial conditions adopted here and during the integration interval considered.

The evolution of orbital inclination $I$ shows smooth variations around 0.136$^\circ$. The cycles may display gravitational perturbation with Ophelia and Bianca, which have similar inclinations. 

The orbital angles $\omega$, $\Omega$, and $M$ exhibit the expected circulation throughout the entire integration interval. No interruptions, long-term confinement episodes, or abrupt changes in their evolution are observed. Together with the small variations in semimajor axis $a$ and eccentricity $e$, this behavior is consistent with a regular orbital evolution of S/2025~U1 over the time span considered.

\section{Hypothetical undiscovered satellites}
\label{sec:hypsat}
\begin{figure*}
    \centering
    \includegraphics[width=\linewidth]{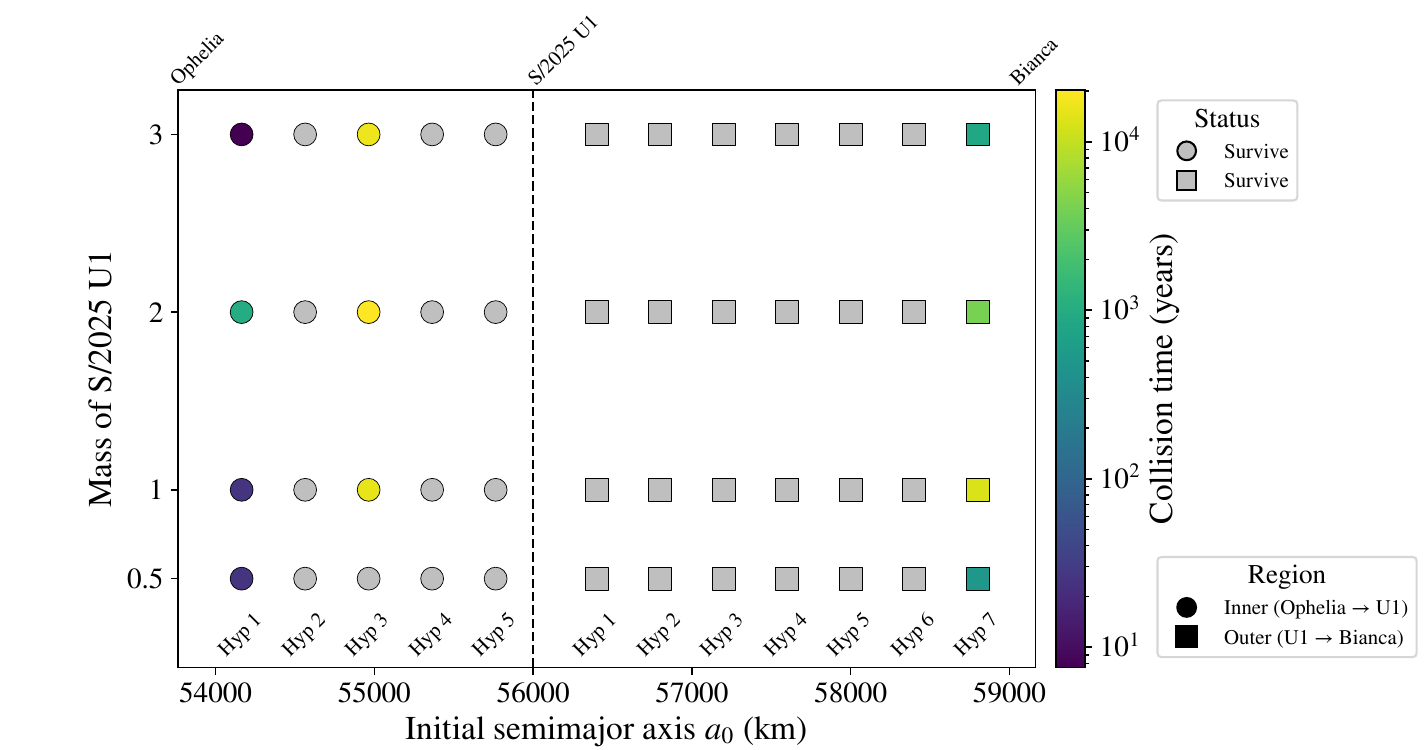}
    \caption{Collision map for 12 hypothetical satellites with masses ranging between 0.5 and three times the mass of S/2025~U1. The circles correspond to hypothetical satellites distributed between the orbits of Ophelia and S/2025~U1 (inner region), while the squares represent hypothetical satellites distributed between the orbits of S/2025~U1 and Bianca (outer region). The gray symbols indicate satellites that survived throughout the entire integration time. The colored symbols, according to the color scale, indicate collisions with one of the neighboring moons or another hypothetical satellite, with the color corresponding to the collision time. The initial conditions were defined within the intervals $54,164 \leq a \leq 55,764$~km (inner region), and $56,400 \leq a \leq 58,800$~km (outer region). All satellites were initially placed on circular coplanar orbits, with $\omega=\Omega=M=0^\circ$, and spaced by 400~km. Hyp denotes a hypothetical satellite.}

    \label{fig:hyp_colision}
\end{figure*}

\begin{figure*}[!ht]
    \centering
    \begin{subfigure}{0.58\textwidth}
        \centering
        \includegraphics[width=\linewidth]{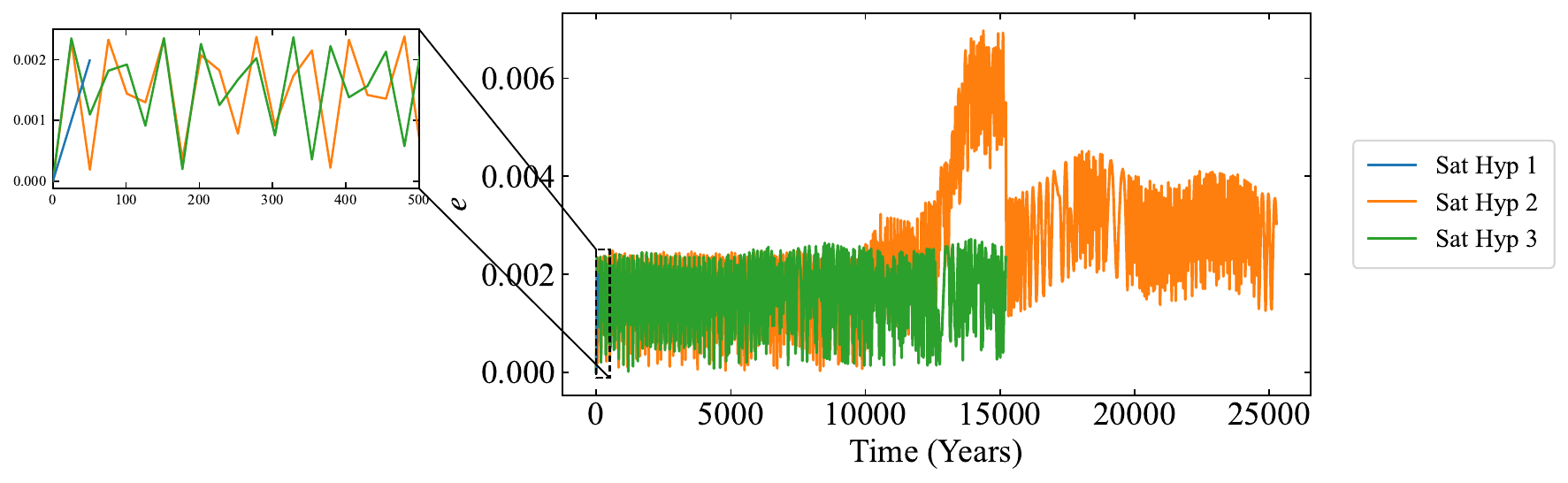}
        \caption{}
            \end{subfigure}
    \hfill
    \begin{subfigure}{0.40\textwidth}
        \centering
        \includegraphics[width=\linewidth]{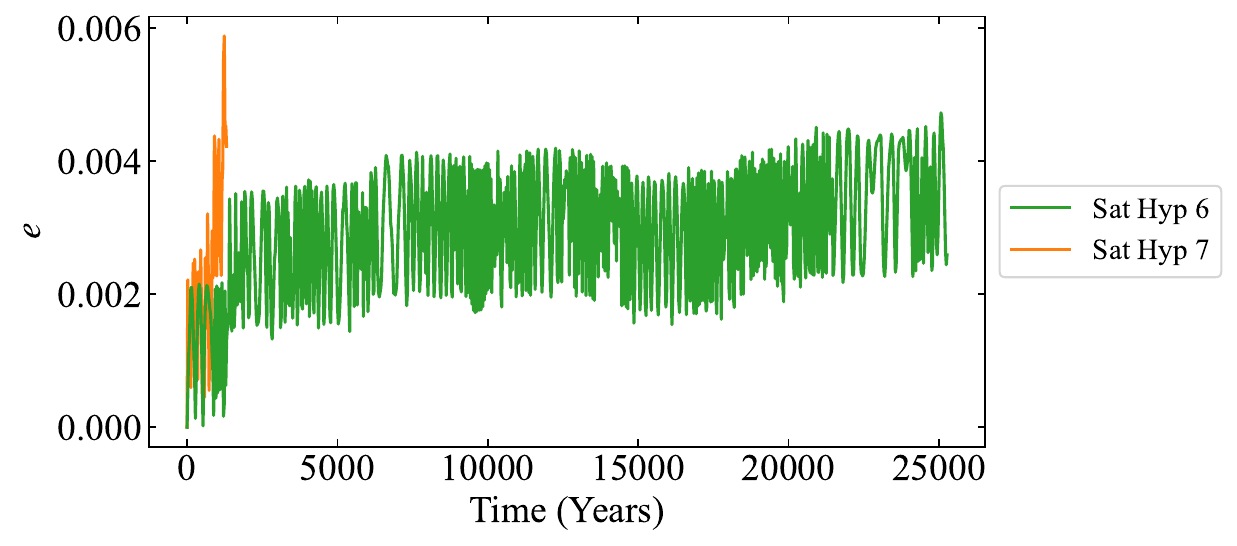}
        \caption{}
   \end{subfigure}
\caption{Temporal evolution of the eccentricity of selected hypothetical satellites. \textit{Panel (a)}: Hypothetical satellites 1, 2, and 3 in the inner region. \textit{Panel (b)}: Hypothetical satellites 6 and 7 in the outer region. In this case, all hypothetical satellites have masses equal to that of S/2025~U1. In the inner region, hypothetical satellite 1 collides with Ophelia after only a few days of integration. Hypothetical satellites 2 and 3 collide with each other, but only satellite 3 did not survive until the end of the integration. In the outer region, hypothetical satellites 6 and 7 also collide; however, only satellite 7 is removed from the system, while satellite 6 remains until the end of the integration.}

    \label{fig:sats_inner_outer}
\end{figure*}
\begin{figure*}[!ht]
\centering
\begin{minipage}{0.49\textwidth}
    \centering
    \includegraphics[width=\linewidth]{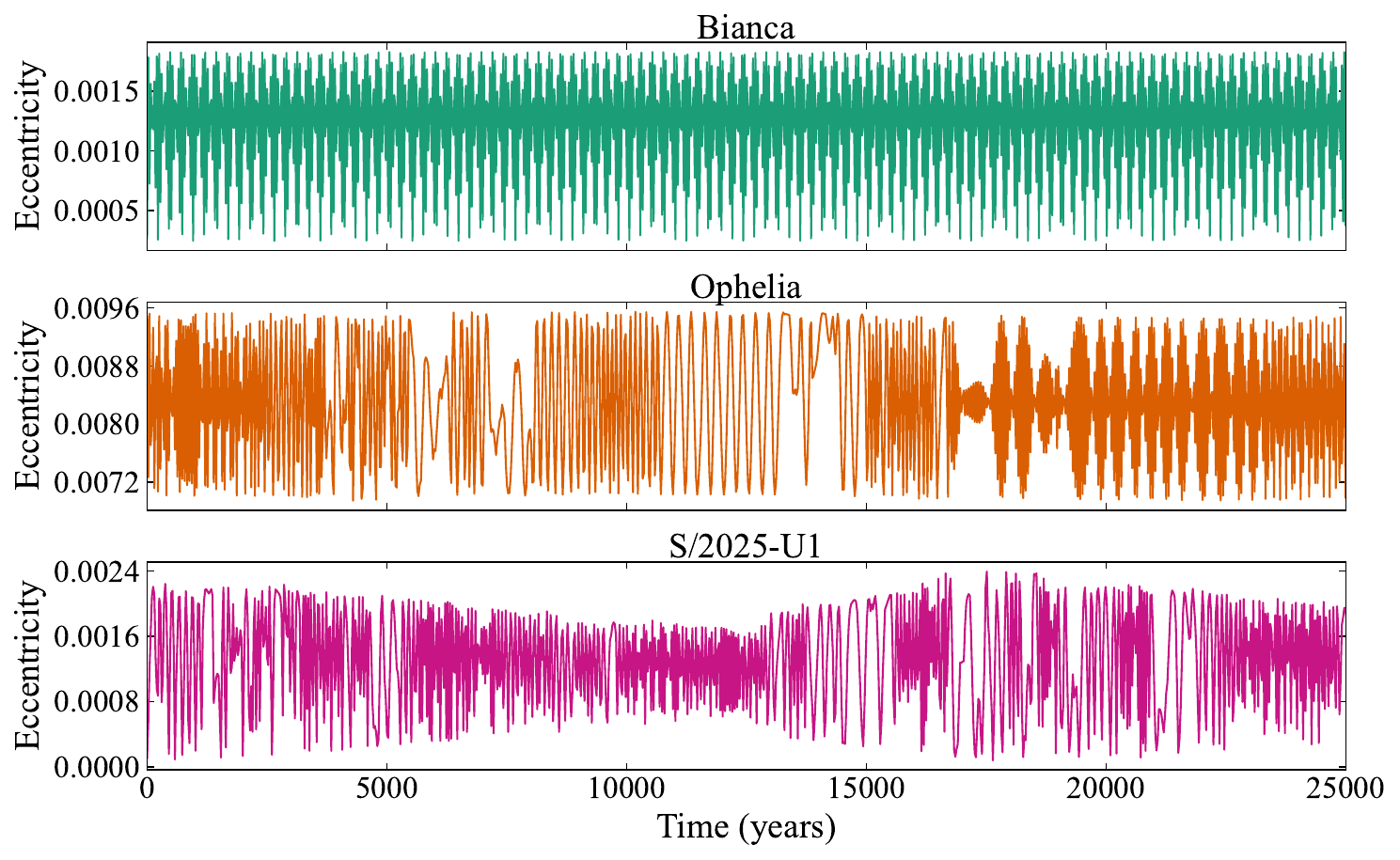}
    \caption*{(a) }
\end{minipage}
\hfill
\begin{minipage}{0.49\textwidth}
    \centering
    \includegraphics[width=\linewidth]{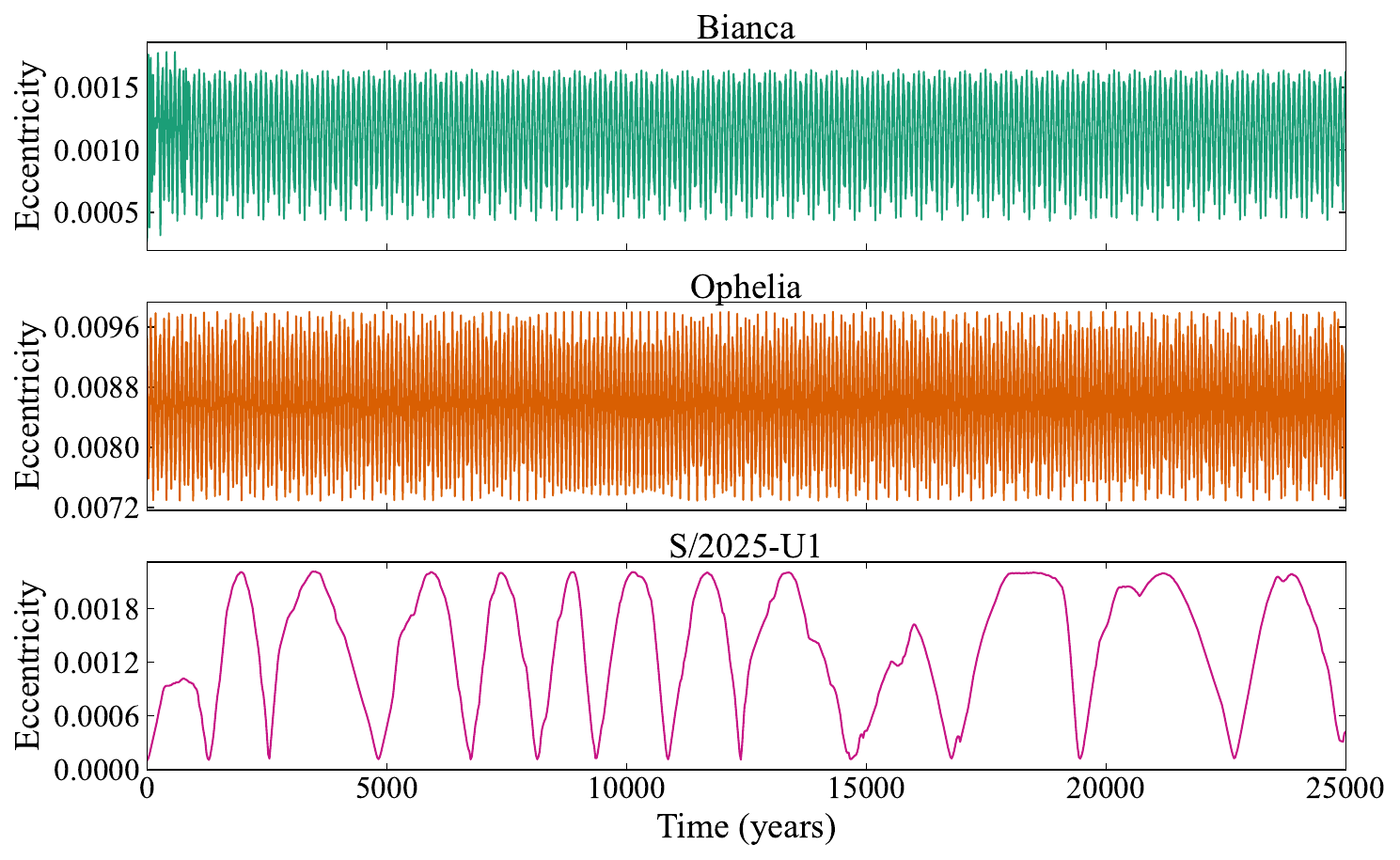}
    \caption*{(b) }
\end{minipage}
\caption{Temporal evolution of the eccentricity of Bianca, Ophelia, and S/2025~U1 for integrations performed with hypothetical satellites with masses equal to three times the mass of S/2025~U1. \textit{Panel (a)}: Temporal evolution of these moons under the influence of hypothetical satellites located in the inner region. \textit{Panel (b)}: Evolution under the influence of hypothetical satellites from the outer region. In these simulations, we adopted a radius of 5 km for S/2025~U1 and the updated mass of Ophelia (New Mass).}
\label{fig:eccentricity_moons_3x}
\end{figure*}
In Sect. \ref{sec:global_difusion} we show that, in the region occupied by S/2025~U1, a satellite with a radius smaller than $\sim 20$ km could survive in orbit without producing significant perturbations on the orbits of Ophelia and Bianca. However, the dynamical space between these three satellites may still accommodate additional bodies of different sizes and masses.

Previous studies of the Pluto-Charon system, for instance, have shown that the dynamical structure of both internal and external regions can be influenced by the presence of satellites, delimiting stable and unstable zones for particles and additional bodies \citep{Stern1994, Pires2011}.

In this section we investigate the possibility of the existence of additional satellites in the regions internal and external to the newly discovered moon S/2025 U1. To this end, we fix the radius of S/2025~U1 at 5.0 km and distribute hypothetical satellites across each region, considering mass values ranging from $0.5$ to $3\times$ the mass of S/2025 U1. We also analyze the stability of S/2025~U1 and the neighboring moons Ophelia and Bianca in the presence of the hypothetical satellites.

In the inner region, located between Ophelia and S/2025~U1, the hypothetical satellites were distributed over a semimajor axis interval ranging from 54,164~km (innermost satellite) to 55,764~km (outermost satellite). A total of five satellites were considered, equally spaced at 400 km intervals.

In the outer region, located between the orbits of S/2025~U1 and Bianca, the hypothetical satellites were distributed over a semimajor axis ranging from 56,400~km to 58,800~km. A total of seven satellites were considered, equally spaced at 400~km intervals. The larger number of satellites in this region compared to the inner region is a consequence of its greater extent. All satellites were initially placed on circular and coplanar orbits, with all remaining orbital elements set to zero. The orbital evolution was then integrated over 25,000 years.

The orbital evolution and stability of the hypothetical undiscovered satellites were analyzed through the identification of collisions, both among the hypothetical satellites themselves and between them and the system's satellites, as well as through the evaluation of the maximum eccentricity $e_{max}$ achieved during the numerical integrations. Satellites that did not experience collisions during the integration time were classified as survivors. In cases where collisions occurred but at least one of the bodies involved remained in orbit until the end of the integration, that satellite was also considered a survivor from an orbital standpoint. Additionally, satellites with maximum eccentricity values $e_{max}<10^{-3}$ were classified as dynamically weakly excited over the analyzed integration time.

Figure \ref{fig:hyp_colision} shows the collision map for the 12 hypothetical satellites with masses ranging between 0.5 and 3 times the mass of S/2025~U1. The circles represent hypothetical satellites distributed in the inner region, between Ophelia and S/2025~U1, while the squares represent satellites in the outer region, between S/2025~U1 and Bianca. The gray symbols indicate satellites that survived the entire integration time, whereas the colored symbols represent satellites that experienced collisions, with the color indicating the collision time.

In the inner region, it is observed that hypothetical satellite 1 (Hyp~1) collides with Ophelia after only a few years of integration for all mass values considered. This result indicates that proximity to Ophelia's orbit makes this region dynamically unstable, independent of the mass of the hypothetical satellite.
Figure \ref{fig:sats_inner_outer}a shows that this collision occurs after approximately 50 years.

Still in the inner region, it is found that hypothetical satellite 3 (Hyp~3) does not survive for masses equal to or greater than the mass of S/2025~U1.
Figure \ref{fig:sats_inner_outer}a shows that, around 13,000 to 15,000 years, a significant increase in eccentricity occurs associated with the encounter between hypothetical satellites 2 and 3. This encounter leads to a collision event at $t \sim 13,000$ years, after which only hypothetical satellite 2 remains in the system and completes the entire integration. Despite the observed perturbation, hypothetical satellite 2 continues to exhibit eccentricities of the order of $10^{-3}$ until the end of the integration, indicating that its orbit remains only moderately excited.

In the region external to the orbit of S/2025~U1, within the adopted semimajor axis interval, between 56,400~km and 58,800~km, the only collision observed in the integrations was that of hypothetical satellite~7 (Hyp~7), for all mass values considered. This collision occurs with hypothetical satellite~6 (Hyp~6). However, only Hyp~7 is removed from the integration, whereas Hyp~6, which despite exhibiting an abrupt increase in eccentricity, remains in the system until the end of the integration, with maximum eccentricities of the order of $10^{-3}$, as shown in Fig. \ref{fig:sats_inner_outer}b.

The results indicate that additional satellites could exist in both the inner and outer regions. However, the survival of the hypothetical satellites depends strongly on their orbital location. In the inner region, satellites closer to Ophelia's orbit are more susceptible to collisions and removal from the system. In contrast, the external region contains a larger number of surviving satellites and fewer collisions, suggesting more favorable conditions for the long-term persistence of additional bodies. These results suggest that the region surrounding S/2025~U1 may still host additional satellites.

However, it is important to emphasize that all satellites were initially assumed to be on circular and coplanar orbits, with the remaining orbital elements set to zero. Therefore, the effects associated with variations in the orbital elements, such as the longitude of the ascending node, the argument of pericenter, and the mean anomaly, were not explored. The inclusion of these parameters may alter the relative configuration of the satellites, influencing the occurrence of close encounters and the overall stability of the system \citep{Duncan1989,Gladman1993}.

In addition, it should be considered that only a limited set of hypothetical satellites and mass values was investigated. Likewise, the region occupied by Uranus's inner satellites presents a complex dynamical structure, including several mean-motion resonances and resonant chains that were not analyzed in this section. Thus, although the results indicate the possibility of orbital stability in both regions, a broader exploration of the $a \times e$ phase space, as well as a detailed investigation of the resonant mechanisms in the system, is necessary to understand these orbital stability conditions better.

The analysis of the hypothetical satellites showed that stable regions may exist between the orbits of the moons neighboring S/2025~U1. However, the simple survival of these bodies does not guarantee that their presence is dynamically compatible with the current system. Thus, it is necessary to investigate whether additional satellites could produce significant perturbations on the orbits of the neighboring moons, especially Ophelia, Bianca, and S/2025~U1 itself.

Figure \ref{fig:eccentricity_moons_3x} shows the temporal evolution of the eccentricity of Bianca, Ophelia, and S/2025~U1 for the case in which the hypothetical satellites have masses equal to three times the mass of S/2025~U1. The choice of this mass value was motivated by the interest in analyzing the influence of the most massive hypothetical satellites considered in this work. The integrations were performed simultaneously for all hypothetical satellites and the three moons under analysis, with all bodies interacting gravitationally throughout the integration. Panel (a) corresponds to the case in which the hypothetical satellites were inserted in the inner region, between Ophelia and S/2025~U1, whereas panel (b) corresponds to the case of the outer region, between S/2025~U1 and Bianca.

In the case of the inner region, it is observed that the eccentricity of Ophelia exhibits more evident modulations throughout the integration, especially after approximately $13,000$ years. This behavior suggests that the hypothetical satellites located between Ophelia and S/2025~U1 produce more noticeable perturbations on Ophelia's orbit. Nevertheless, the eccentricity remains limited to values of the order of $10^{-3}$, indicating that the perturbation does not lead to significant secular growth during the considered integration interval.

On the other hand, in the case of the outer region, the evolution of Ophelia remains more regular throughout the integration. Bianca also exhibits small eccentricity variations in both cases, without abrupt changes associated with the hypothetical satellites. This result indicates that Bianca is only weakly affected by the analyzed configurations, even in the case where the hypothetical satellites are placed in the outer region.

The moon S/2025~U1 exhibits the largest relative eccentricity variations among the three analyzed bodies. This behavior is expected, since S/2025~U1 is located between the two regions where the hypothetical satellites were distributed. In the inner case, its evolution shows more irregular variations, particularly around $\sim 13,000$ years. This time interval is close to the instant of the collision between Hyp~2 and Hyp~3, suggesting that the removal of one of the hypothetical satellites may be associated with the observed change in the orbital evolution of S/2025~U1. In contrast, in the outer case, the evolution remains smoother. Thus, the results suggest that the inner region produces a more significant perturbation on the local system than the outer region.

In general, the integrations indicate that the presence of the hypothetical satellites does not produce large changes in the orbits of Bianca, Ophelia, and S/2025~U1 over $25,000$ years. However, the comparison between the panels shows that the satellites in the inner region exert a more noticeable dynamical influence, especially on Ophelia and S/2025~U1. In contrast, the outer region produces weaker, more regular perturbations.

\section{Conclusions}
\label{sec:concl}
In this work, we have performed an analysis of the orbital stability of the inner satellites of Uranus incorporating three significant updates: (i) the inclusion of the newly discovered moon S/2025 U1; (ii) revised mass estimates for Cordelia, Ophelia, and Cressida; and (iii) updated Uranian zonal harmonic coefficients. Through numerical integrations, mean-motion resonance analysis, and diffusion maps, we investigated how these updates affect the orbital evolution and stability of the system. We also, for the first time, explored the dynamical environment of the newly discovered moon S/2025~U1.

The analysis based on the diffusion parameter, using the FMA technique to evaluate the orbital stability of Uranus's inner satellites under the proposed updates, revealed that small variations in satellite masses and certain physical parameters can locally modify the dynamics in densely populated regions of the system, such as the subgroup composed of Bianca, Cressida, and Desdemona. These satellites exhibit greater sensitivity to such modifications, as evidenced by changes in the diffusion parameter $D$. Similarly, when analyzing variations in the maximum eccentricity $e_{max}$, this group shows more significant changes than the other satellites under these updates.

By analyzing the temporal evolution of the resonant angles to investigate resonant confinement among the satellites, we find that most resonances involving pairs of satellites exhibit a circulating behavior. However, the 44:43 resonance between Belinda and Perdita exhibits low-amplitude libration throughout the entire integration. Although this pair is located in a narrow region of orbital instability, the 44:43 resonance may contribute to the dynamical stability of these satellites. This region, located between Cupid and Puck (R6 and R7), also corresponds to an area where we identified a significant concentration of first-order resonances involving test particles and the satellites.

We analyzed the dynamics of the newly discovered moon S/2025~U1 and its surrounding region. We investigated how variations in the radius of a hypothetical satellite in the orbit of the new moon affect the local dynamics. We showed that even a satellite with a radius of only a few kilometers can influence the stability of nearby particles.

Diffusion maps indicate that the radius most consistent with stable dynamic behavior is approximately 5 km, in agreement with photometric estimates from JWST. For this value, the moon remains stable for more than 250,000 years, exhibiting a smooth orbital evolution without evidence of significant long-term variations.

However, our results show that for $R = 7$ km, a reduction in the diffusion of neighboring satellites, such as Bianca, is observed, suggesting a local stabilizing effect for the orbital evolution of these satellites. Taken together, these results indicate that the stability of this group may be associated with radius values in the range $5 \leq R \leq 7$ km.

Our simulations also indicate that a satellite with a radius equal to ~20 km would produce perceptible gravitational perturbations in the orbits of adjacent satellites, primarily Desdemona. This behavior is incompatible with the current dynamic orbital configuration of the system.

The diffusion map results suggest that the vicinity of S/2025~U1 contains dynamically stable regions capable of hosting particles and possibly even small satellites. To investigate these stability regions both inside and outside the orbit of the new moon, we performed simulations with 12 hypothetical satellites, each with a mass ranging from 0.5 to three times that of S/2025~U1. We observed that the region between Ophelia and S/2025~U1 (inner region) is dynamically more unstable since the hypothetical satellite closest to Ophelia's orbit collides with this moon for all mass values considered after a short integration time. In addition, we identified a second collision event involving hypothetical satellites 2 (Hyp~2) and 3 (Hyp~3) for masses equal to or greater than the mass of S/2025~U1. The remaining satellites remained stable throughout the integrations, exhibiting maximum eccentricities below $10^{-3}$. Even in the most massive scenarios, corresponding to three times the mass of S/2025~U1, the surviving satellites remained stable for up to 25,000 years, producing only moderate variations in the orbits of the neighboring moons Ophelia and Bianca.

Our results also showed that the region between the orbit of S/2025~U1 and Bianca (outer region) appears to be dynamically more stable. In this region, only one collision event was observed, and it involved the hypothetical satellites 6 (Hyp~6) and 7 (Hyp~7). The remaining satellites survived the integration, with maximum eccentricities below $10^{-3}$.

Finally, this work presents the first detailed analysis of the dynamics of the newly discovered moon S/2025~U1 as well as the first study of Uranus's inner satellite system incorporating the most recent updates to its physical parameters. Our results not only refine the understanding of orbital stability in this highly compact region but also identify dynamically stable regions in the vicinity of the new moon that could potentially host additional small satellites, opening new perspectives for future studies.

\begin{acknowledgements}
A.~Amarante and A.~Ferreira thank the financial support of the São Paulo Research Foundation (FAPESP) [grants \#2023/11781-5 \& \#2025/15438-9]. SGW thanks FAPESP (Proc~2022/11783-5) and CNPq (Proc.~309057/2025-6) for the financial support. A. Amarante thanks T. Gallardo for the discussion at the XIII Taller CP 2026 on the mean-motion resonances among Uranus's inner satellites.
The authors also acknowledge the financial support of the Coordination for the Improvement of Higher Education Personnel (CAPES) - Brazil (finance code 001). This research was also supported by computational resources supplied by the Center for Scientific Computing (NCC/GridUNESP) of the São Paulo State University (UNESP) and the Center for Mathematical Sciences Applied to Industry (CeMEAI), funded by FAPESP [grant \#2013/07375-0].
In preparing this work, the authors utilized Grammarly to enhance the readability of the English. Following this, the authors carefully reviewed and edited the content, assuming full responsibility for the publication's material.
\end{acknowledgements}

\bibliographystyle{aa}
\bibliography{references}

\end{document}